\documentclass[%
 preprint, 
nofootinbib,
 amsmath,amssymb,
 aps, physrev,
]{revtex4-2}

\usepackage{amsmath, amssymb, amsfonts}
\allowdisplaybreaks
\usepackage{dsfont}
\usepackage{graphicx}
\usepackage{bm}
\usepackage{hyperref}

\usepackage{braket}

\usepackage{xcolor}

\begin{document}
\title{Analytical Landscape of Maximal Magic for Two-Qutrit States and Beyond}
\author{Marco Knipfer}
\author{Alexander Roman}
\author{Katia Matcheva}
\author{Konstantin T. Matchev}
\affiliation{%
	Department of Physics and Astronomy, University of Alabama, Tuscaloosa, AL USA
}%
\date{\today}
\begin{abstract}
	Achieving a genuine quantum advantage relies on two distinct non-classical resources that restrict efficient classical simulation: entanglement and magic (nonstabilizerness). We investigate the interplay between these resources by characterizing the Pareto frontiers of extreme magic at fixed entanglement for systems of two qutrits ($d=3$) and two ququints ($d=5$). Unlike the case of two qubits, the Schmidt spectrum for two qutrits features two independent entanglement parameters, resulting in two-dimensional Pareto surfaces. For the lower frontier, we recast the minimal magic as a compact function of concurrence and negativity, with a maximal value of $\ln 2$. For the upper frontier, we determine the maximal stabilizer R\'enyi entropy to be $M_2 = \ln(81/17) \approx 1.561$, which tightens the previous theoretical bound of $\ln 5\approx 1.609$ and improves on earlier numerical estimates. The maximum magic is achieved at eighteen distinct maxima categorized into three families of six permutation-equivalent spectra. We provide analytical expressions for the maximal magic in the neighborhood of each maximum and for the corresponding maximally magical states which turn out to be Weyl–Heisenberg-covariant fiducial states for mutually unbiased bases. Finally, numerical analysis of two ququints ($d=5$) reveals six permutation-inequivalent maxima with a peak magic value of $M_2 = \ln(625/49) \approx 2.546$. Based on these findings, we conjecture that the maximal magic for a bipartite system of two qudits with prime dimension $d$ is given by $\ln [ d^4 / (2d^2 - 1) ]$, which reproduces the previously known value for qubits, as well as the values  derived here for qutrits and ququints.
\end{abstract}

\maketitle

\tableofcontents

\section{Introduction}
\label{sec:intro}
The difficulty of simulating a quantum system on classical hardware is governed by various quantum resources, including entanglement and magic \cite{Chitambar:2018rnj}.
States with low entanglement admit an efficient classical description in terms of tensor networks, irrespective of their magic content.
Conversely, by the Gottesman--Knill theorem \cite{Gottesman:1998hu,Gottesman:1999tea,Aaronson:2004xuh}, states with low magic can be simulated efficiently with stabilizer techniques, irrespective of their entanglement.
This observation led Bravyi and Kitaev to promote magic to a quantum resource in its own right, which quantifies the overhead of classical simulation \cite{Bravyi:2004isx}.
Genuine quantum advantage therefore requires states that are simultaneously rich in \textit{both} resources.
Rather than attempting to survey the large and rapidly growing literature on entanglement and magic as quantum resources, we refer the reader to the recent comprehensive review~\cite{Robin:2026lqp}.

While each resource is by now well understood in isolation, their interplay is much less explored \cite{Iannotti:2025lkb,Qian:2025oit,Liu:2025frx,Mittal:2025gnd,Busoni:2026lvp,Roman:2026mcy}.
The two resources are not independent:
part of the magic of a state resides in the local degrees of freedom, where it can be freely created or removed by local unitary operations, while the remainder, the \textit{non-local magic}, is inseparably linked to entanglement~\cite{Robin:2026lqp}.
As a consequence, the region of the entanglement--magic parameter space that is actually populated by physical states has a non-trivial shape, delineated by the boundaries of maximal and minimal magic, respectively.
Recent work in Ref.~\cite{Roman:2026mcy} referred to those boundaries as the \textit{Pareto frontiers} of magic and entanglement and derived them analytically for the simplest non-trivial system, that of two qubits.
As is customary, the entanglement was quantified by the concurrence $\Delta$ \cite{Hill:1997pfa,Wootters:1997id} and the magic by the stabilizer R\'enyi entropy of order two, $M_2$~\cite{Leone:2021rzd}. For a two-qubit system, the frontier of minimal magic, $M_2^{(\text{min})}(\Delta)$, is a single continuous curve, whose analytic form was previously derived in \cite{Qian:2025oit,Busoni:2026lvp}. On the other hand, the frontier of maximal magic, $M_2^{(\text{max})}(\Delta)$, turned out to be quite complex: it is piecewise defined in terms of three separate analytic segments \cite{Roman:2026mcy}. The maximal value of $M_2$ is achieved at two different locations, $\Delta=\frac{1}{2}$ and $\Delta=\frac{1}{\sqrt{2}}$ \cite{Liu:2025frx}.

A related issue often discussed in the literature is the question of the {\em maximal possible magic} in a system of two qudits \cite{Wang:2023uog,Cuffaro:2024wet,Liu:2025frx,Ohta:2025utz,Erew:2025egp}. (A qudit is the $d$-level generalization of a qubit, i.e., a quantum system whose Hilbert space is~$\mathbb{C}^d$.) Refs.~\cite{Wang:2023uog,Cuffaro:2024wet} derived an upper bound on the maximal value of $M_2$ in a two qudit system of
\begin{equation}
	\max(M_2) \le \ln\frac{d^2}{1+\frac{d^2-1}{d^2+1}},
	\label{eq:Cuffaro_bound}
\end{equation}
where $d^2$ is the dimensionality of the Hilbert space of the two-qudit system. For qubits $d=2$ and eq.~(\ref{eq:Cuffaro_bound}) predicts a maximal value of $\ln (5/2) \approx 0.916$. This, however, turned out to be an overestimate, and a tighter bound of $\ln (16/7)\approx 0.827$ was found in Ref.~\cite{Liu:2025frx} and confirmed with the analytical expressions for $M_2^{(\text{max})}(\Delta)$ found in \cite{Roman:2026mcy}.

In this paper, we take the next step and study the Pareto frontiers of magic and entanglement for a system of two qudits, working out in detail the case of two qutrits.
Qutrits are not only of practical interest as a computational platform in their own right~\cite{Ringbauer:2021lhi, PhysRevA.58.883}.
From the perspective of the Pareto frontiers, the step from qubits to qutrits is also a qualitative one.
For two qubits, the Schmidt decomposition contains a single independent parameter, which is in one-to-one correspondence with the concurrence $\Delta$, so that a single scalar labels the orbits of local unitary transformations.
For two qutrits, in contrast, the Schmidt spectrum~$\boldsymbol{\lambda}\equiv (\lambda_0,\lambda_1,\lambda_2)$ carries two independent parameters (after taking into account the normalization constraint).
Therefore, no single scalar measure of entanglement can resolve the local-unitary orbits, and the natural setting for the analysis is the two-dimensional simplex of Schmidt coefficients, which can be visualized with, e.g., ternary plots.
Accordingly, the one-dimensional Pareto frontiers of the two-qubit problem considered in \cite{Roman:2026mcy} are now promoted to two surfaces over the Schmidt simplex:
the minimal magic $M_2^{(\mathrm{min})}(\boldsymbol{\lambda})$ and the
maximal magic $M_2^{(\mathrm{max})}(\boldsymbol{\lambda})$.
Closed-form expressions for the minimal magic $M_2^{(\mathrm{min})}(\boldsymbol{\lambda})$ of two-qutrit systems were recently obtained in~\cite{Busoni:2026lvp}, so here we shall focus on the case of maximal magic, $M_2^{(\mathrm{max})}(\boldsymbol{\lambda})$.

Our main results and the organization of the paper are as follows:
\begin{itemize}
	\item After introducing our conventions and notation in Section~\ref{sec:notation}, in Section~\ref{sec:minmagic} as a warm-up we rederive the result from \cite{Busoni:2026lvp} for the minimal magic surface $M_2^{(\mathrm{min})}(\boldsymbol{\lambda})$. We then suggest a simpler equivalent parametrization $M_2^{(\mathrm{min})}(C,N)$ in terms of two entanglement measures, the I-concurrence $C$ and the negativity $N$ defined in Section~\ref{sec:entanglement}.
	\item Our main results for $M_2^{(\mathrm{max})}(\boldsymbol{\lambda})$ are presented in Section~\ref{sec:maxmagic}. We show that the maximal magic is achieved at 18 different locations in the two-dimensional simplex of Schmidt coefficients. These 18 maxima are degenerate and can be grouped into 3 distinct groups of 6. Within each group, the $(\lambda_0,\lambda_1,\lambda_2)$ values at the maxima are related by permutations. Each set of $\lambda$'s satisfies a cubic equation whose coefficients depend on the respective entanglement parameters $C$ and $G$.
	\item We find that the maximal magic for a system of two qutrits is given by $\ln(81/17)\approx 1.561$, which tightens the prediction of $\ln 5\approx 1.609$ given by eq.~(\ref{eq:Cuffaro_bound}). Our result also improves the previous numerically derived result of $1.548$ quoted in Ref.~\cite{Chernyshev:2024pqy}.
	\item We obtain analytical functions for $M_2^{(\mathrm{max})}(\boldsymbol{\lambda})$ which are exact in the neighborhood of each maximum. The validity of these analytical expressions over the entire simplex (far away from the maxima) is studied in Section~\ref{sec:validity}.
	\item In Section~\ref{sec:generalization}, we generalize our discussion to the case of two qudits and propose that the exact bound that should replace eq.~(\ref{eq:Cuffaro_bound}) is
	      \begin{equation}
		      \max(M_2) \le \ln\frac{d^2}{1+\frac{d^2-1}{d^2}} = \ln \frac{d^4}{2d^2-1}.
		      \label{eq:Our_bound}
	      \end{equation}
	      This new and improved bound agrees with a) the previous result of $\ln (16/7)$ for qubits \cite{Liu:2025frx}; b) the value of
	      $\ln (81/17)$ for qutrits derived in Section~\ref{sec:maxmagic}; and c) the answer of $\ln (625/49)$ for a system of two ququints ($d=5$) derived numerically in Section~\ref{sec:ququints}. In all cases, the states of maximal magic are Weyl–Heisenberg-covariant fiducial states for mutually unbiased bases, just like in the qubit case \cite{Liu:2025frx}.
\end{itemize}
We summarize our results and discuss their implications in Section~\ref{sec:conclusions}. Details (and some subtleties) of the numerical optimization procedure are presented in Appendix~\ref{app:numerics}.

\section{Conventions and Notations for Two-Qutrit States}
\label{sec:notation}

\subsection{Two-Qutrit State Parametrization}

A general state $\vert\psi\rangle$ of two qutrits is a superposition of the nine computational basis states $\vert ij\rangle$, $i,j=0,1,2$:
\begin{equation}
	\vert \psi \rangle =
	\sum_{i,j=0}^2 a_{ij} \vert ij \rangle \,,
	\label{eq:generic_psi_def}
\end{equation}
with some complex coefficients $a_{ij}$ normalized as
\begin{equation}
	\sum_{i,j=0}^2 |a_{ij}|^2=1.
	\label{eq:normalization}
\end{equation}
Note that in this paper we use zero-based indexing for the computational basis states. Eliminating a global phase and accounting for the constraint (\ref{eq:normalization}) reduces the 2-qutrit parameter space from an 18-dimensional to a 16-dimensional manifold.
For our purposes here, it is sufficient to parametrize this manifold in terms of the Schmidt decomposition
\begin{equation}
	\ket{\psi(\boldsymbol{\varphi}_A, \boldsymbol{\varphi}_B, \boldsymbol{\lambda})} =
	\left(U_A(\boldsymbol{\varphi}_A) \otimes U_B(\boldsymbol{\varphi}_B)\right)\ket{\psi_{\mathrm{S}}(\boldsymbol{\lambda})},
	\label{eq:Schmidt_decomposition_LU}
\end{equation}
where $U_A$ and $U_B$ are $3\times 3$ unitary matrices
and therefore their tensor product $U_A\otimes U_B$ is a $9\times 9$ matrix acting on the 9-dimensional Schmidt reference state
\begin{equation}
	\ket{\psi_{\mathrm{S}}(\boldsymbol{\lambda})}
	\equiv
	\sum_{i=0}^2 \sqrt{\lambda_i}\,\ket{ii}
	=
	\left(
	\begin{array}{c}
			\sqrt{\lambda_0} \\
			0                \\ 0 \\ 0 \\
			\sqrt{\lambda_1} \\
			0                \\ 0 \\ 0 \\
			\sqrt{\lambda_2}
		\end{array}
	\right).
	\label{eq:def_Schmidt_state}
\end{equation}
The Schmidt reference state $\ket{\psi_{\mathrm{S}}(\boldsymbol{\lambda})}$ is parametrized in terms of three real and non-negative numbers
\begin{equation}
	\boldsymbol{\lambda} \equiv (\lambda_0,\lambda_1,\lambda_2),
	\qquad
	\lambda_i \ge 0,
\end{equation}
which we shall refer to as Schmidt coefficients.
The unit-normalization of $\ket{\psi_{\mathrm{S}}(\boldsymbol{\lambda})}$ implies that the Schmidt coefficients $\boldsymbol{\lambda}$ obey the following normalization condition
\begin{equation}
	\lambda_0+\lambda_1+\lambda_2 = 1.
	\label{eq:lambda_normalization}
\end{equation}
The Schmidt reference state (\ref{eq:def_Schmidt_state}) therefore depends on only 2 real degrees of freedom. The remaining 14 degrees of freedom parametrizing the general state (\ref{eq:Schmidt_decomposition_LU}) reside in the parameters $\boldsymbol{\varphi}_A$ and $\boldsymbol{\varphi}_B$ of the unitary matrices $U_A$ and $U_B$.

We should point out that there are alternative conventions for the Schmidt coefficients in the literature, e.g., $\ket{\psi_{\mathrm{S}}(\boldsymbol{\kappa})}
	\equiv
	\sum_{i=0}^2 \kappa_i\,\ket{ii}$, with a normalization condition $\kappa_0^2+\kappa_1^2+\kappa_2^2=1$ in place of (\ref{eq:lambda_normalization}). Our conventions are fixed by eqs.~(\ref{eq:def_Schmidt_state}) and (\ref{eq:lambda_normalization}), which allows us to visualize the $\boldsymbol{\lambda}$ parameter space with a ternary plot that automatically incorporates the normalization constraint (\ref{eq:lambda_normalization}), see Section~\ref{sec:plotting} below.

\subsection{Measures of Entanglement}
\label{sec:entanglement}
The entanglement structure of a pure bipartite quantum state is encoded in its Schmidt coefficients. For separable states, there is only one non-vanishing coefficient, while for maximally entangled states, all coefficients are equal: $\lambda_i = \frac{1}{d}$, $\forall i \in \{0, 1, \ldots, d-1\}$. For a two-qubit state, there are only two Schmidt coefficients, $\lambda_0$ and $\lambda_1$, which are constrained by the $d=2$ analogue of eq.~(\ref{eq:lambda_normalization}), namely  $\lambda_0+\lambda_1=1$. As a result, the entanglement of two qubits can be quantified by a single degree of freedom, for which there are different equivalent choices in the literature, including the concurrence $\Delta(\vert \psi\rangle)=2\sqrt{\lambda_0\lambda_1}$ \cite{Hill:1997pfa,Wootters:1997id}, where the factor of two is chosen so that the maximum value of $\Delta$ is normalized to 1. Previous analyses of the relation between magic and entanglement in two qubit systems used the concurrence $\Delta$ as a measure of entanglement \cite{Liu:2025frx,Iannotti:2025lkb,Roman:2026mcy}.

The case of two qutrits, however, is a bit more involved. There are three Schmidt coefficients and a single constraint (\ref{eq:lambda_normalization}), which leaves {\em two} degrees of freedom. One can therefore define two different and inequivalent measures of entanglement. For those we shall only consider quantities which are symmetric under permutations of $\lambda_0$, $\lambda_1$, and $\lambda_2$. First, in analogy to $\Delta$, one can define the so-called $G$-concurrence in terms of the rescaled geometric mean of the $\lambda$s as \cite{Gour:2005cys,Cappellini:2006eot}:
\begin{equation}
	G \equiv 3 \left(\lambda_0 \lambda_1 \lambda_2\right)^{\frac{1}{3}} ,
	\label{eq:G_def}
\end{equation}
where the prefactor of 3 ensures that the maximal value of $G$ is again equal to 1.\footnote{When using the alternative notation for the Schmidt coefficients, $\kappa_i=\sqrt{\lambda_i}$, one can define a different version of G-concurrence as $G_k\equiv 3 \left(\kappa_0 \kappa_1 \kappa_2\right)^{\frac{1}{3}}$ which is related to our definition in (\ref{eq:G_def}) as $G_k^2 = 3G$.}

The second independent measure of entanglement can be taken to be the I-concurrence~\cite{Rungta:2001zcj}
\begin{equation}
	C \equiv \left( \frac{3}{2}\,
	\Bigl[1-\Bigl(\lambda_0^2 + \lambda_1^2 +\lambda_2^2\Bigr)  \Bigr]\right)^\frac{1}{2},
	\label{eq:C_def1}
\end{equation}
where again the factor of $3/2$ under the square root ensures that the maximal value of $C$ is normalized to 1. Using (the square of) the normalization condition (\ref{eq:lambda_normalization}), one can equivalently rewrite (\ref{eq:C_def1}) in terms of $\lambda$ pairings:
\begin{equation}
	C = \left[3\,
		\Bigl(\lambda_0\lambda_1 + \lambda_1\lambda_2 +\lambda_0\lambda_2\Bigr)  \right]^\frac{1}{2}.
	\label{eq:C_def}
\end{equation}

Note that the $\lambda_i$ combinations appearing in eqs.~(\ref{eq:lambda_normalization}), (\ref{eq:G_def}) and (\ref{eq:C_def}) are reminiscent of Vieta's formulas for the three roots of a cubic equation. Indeed, any allowed set of $\lambda_0$, $\lambda_1$ and $\lambda_2$ trivially satisfies the cubic equation for a variable $\lambda$
\begin{equation}
	(\lambda-\lambda_0) (\lambda-\lambda_1) (\lambda-\lambda_2) = 0.
	\label{eq:cubic_product}
\end{equation}
Using eqs.~(\ref{eq:lambda_normalization}), (\ref{eq:G_def}) and (\ref{eq:C_def}) and rescaling to avoid fractional numerical coefficients, this equation can be rewritten as
\begin{equation}
	27 \lambda^3 - 27 \lambda^2 + 9 C^2 \lambda - G^3 =0,
	\label{eq:cubic}
\end{equation}
where $G$ and $C$ are the G-concurrence and I-concurrence, respectively. Equation (\ref{eq:cubic}) allows us to switch back and forth between two different representations of the independent variables in our problem: a) the three degrees of freedom $\lambda_0$, $\lambda_1$ and $\lambda_2$ which are subject to the normalization constraint (\ref{eq:lambda_normalization}); and b) the two independent degrees of freedom $G$ and $C$. This connection will be further discussed and illustrated in Section~\ref{sec:plotting} below.

Another entanglement measure worth mentioning is the negativity $N$~\cite{PhysRevA.58.883, Vidal:2002zz}, which is defined as the sum of the products of the (square roots of the) Schmidt coefficients:
\begin{equation}
	N \equiv
	\sqrt{\lambda_0}\sqrt{\lambda_1}
	+\sqrt{\lambda_1}\sqrt{\lambda_2}
	+\sqrt{\lambda_0}\sqrt{\lambda_2}.
	\label{eq:N_def}
\end{equation}
The negativity is already canonically normalized: the negativity of a product state (a single $\lambda_i$ non-zero) is 0 and the negativity of a maximally entangled state (all $\lambda$'s the same) is equal to $3\, \sqrt{\frac{1}{3}}\, \sqrt{\frac{1}{3}} =1$. In our main discussion below we shall encounter negativity-like structures like the $N_{\pm\pm\pm}$ functions defined in eq.~(\ref{eq:f_def}).

The three entanglement measures are basic symmetric polynomials of the same variables ($\lambda_0, \lambda_1, \lambda_2$) and are related to each other as follows
\begin{equation}
	N^2 = \frac{1}{3}\, C^2 + 2\, \sqrt{1+2N} \,\left(\frac{G}{3}\right)^{3/2}.
	\label{eq:NCG}
\end{equation}

\subsection{Measures of Magic}

For a single-qutrit, the Pauli operators
\begin{equation}
	P_{a,b} = X^a Z^b,
	\qquad a,b \in \mathbb{F}_3 \equiv \{0,1,2\},
\end{equation}
are defined in terms of the shift and clock operators
\begin{equation}
	X =
	\begin{pmatrix}
		0 & 0 & 1 \\
		1 & 0 & 0 \\
		0 & 1 & 0
	\end{pmatrix},
	\qquad
	Z =
	\begin{pmatrix}
		1 & 0      & 0        \\
		0 & \omega & 0        \\
		0 & 0      & \omega^2
	\end{pmatrix},
	\qquad
	\omega = e^{2\pi i/3}.
	\label{eq:shiftclock}
\end{equation}
For a two-qutrit system, the operator basis is therefore
\begin{equation}
	P_{\mathbf{a}}\equiv
	P_{a_1,b_1}\otimes P_{a_2,b_2},
	\qquad \mathbf{a} = (a_1, a_2, b_1, b_2) \in \mathbb{F}_3^4 \equiv \{0,1,2\}^{\otimes 4},
	\label{eq:Padef}
\end{equation}
resulting in \(3^4=81\) Pauli 2-strings in total. The order-2 stabilizer Rényi entropy, also referred to as ``magic'', is given by
\begin{equation}
	M_2(\psi)
	=
	-\ln \Pi_2(\psi),
	\label{eq:M2_def}
\end{equation}
where the stabilizer purity $\Pi_2$ is defined as
\label{eq:def_magic}
\begin{equation}
	\Pi_2(\psi)
	=
	\frac{1}{9}
	\sum_{\mathbf{a}\in \mathbb{F}_3^4}
	\left|
	\bra{\psi}
	P_{\mathbf{a}}
	\ket{\psi}
	\right|^4.
	\label{eq:purity}
\end{equation}

\subsection{Optimization Setup}
\label{sec:setup}
Our main task here is to derive analytical functions which give the maximal magic, $M_2^{(\text{max})}$, for a given level of entanglement, i.e.
\begin{equation}
	M_2^{(\text{max})} (\boldsymbol{\lambda})\equiv
	\max_{\boldsymbol{\varphi}_A, \boldsymbol{\varphi}_B}
	M_2(\psi(\boldsymbol{\varphi}_A, \boldsymbol{\varphi}_B, \boldsymbol{\lambda})),
	\label{eq:M2max_def}
\end{equation}
where the optimization is over the parameters $\boldsymbol{\varphi}_A$ and $\boldsymbol{\varphi}_B$ of the local unitaries $U_A$ and $U_B$. Note that $U_A$ and $U_B$ affect the magic but not the entanglement of the state, which only depends on $\boldsymbol{\lambda}$. In light of the definition (\ref{eq:M2_def}), the task can be equivalently stated as finding the minimal purity:
\begin{equation}
	\Pi_2^{(\text{min})} (\boldsymbol{\lambda})\equiv
	\min_{\boldsymbol{\varphi}_A, \boldsymbol{\varphi}_B}
	\Pi_2(\psi(\boldsymbol{\varphi}_A, \boldsymbol{\varphi}_B, \boldsymbol{\lambda})).
	\label{eq:Pi2min_def}
\end{equation}

Since the entanglement is determined by the values of the Schmidt coefficients $\boldsymbol{\lambda}=(\lambda_0,\lambda_1,\lambda_2)$, the maximal magic $M_2^{(\text{max})}$ and the minimal purity $\Pi_2^{(\text{min})}$ are functions of $\boldsymbol{\lambda}$, as indicated in eqs.~(\ref{eq:M2max_def}) and~(\ref{eq:Pi2min_def}).
According to the Schmidt decomposition in eqs.~(\ref{eq:Schmidt_decomposition_LU}) and~(\ref{eq:def_Schmidt_state}), any general state $\psi$ depends linearly on $\sqrt{\lambda_0}$, $\sqrt{\lambda_1}$,  and $\sqrt{\lambda_2}$.
In turn, each term in the purity sum (\ref{eq:purity}) represents a product of two $\psi$ states, which is then raised to the 4-th power. As a result, the purity $\Pi_2$ in general is given by a homogeneous 8-th degree polynomial in the variables $\lambda_0^{\frac{1}{2}}$, $\lambda_1^{\frac{1}{2}}$,  and $\lambda_2^{\frac{1}{2}}$:
\begin{equation}
	\Pi_2(\psi(\boldsymbol{\varphi}_A,\boldsymbol{\varphi}_B,\lambda_0,\lambda_1,\lambda_2)) =
	\sum_{\alpha=0}^8
	\sum_{\beta=0}^{8-\alpha}
	\sum_{\gamma=0}^8\,
	C_{\alpha\beta\gamma}(\boldsymbol{\varphi}_A,\boldsymbol{\varphi}_B)\,
	\lambda_0^{\frac{\alpha}{2}}\,
	\lambda_1^{\frac{\beta}{2}}\,
	\lambda_2^{\frac{\gamma}{2}}\,
	\delta_{(\gamma,8-\alpha-\beta)},
	\label{eq:Pi2_polynomial}
\end{equation}
where the Kronecker symbol $\delta_{(\gamma,8-\alpha-\beta)}$ eliminates the sum over $\gamma$ and ensures that $\alpha+\beta+\gamma=8$. The minimization in (\ref{eq:Pi2min_def}) fixes the values of $\boldsymbol{\varphi}_A$ and $\boldsymbol{\varphi}_B$, rendering the coefficients $C_{\alpha\beta\gamma}$ into numerical constants. After reorganizing terms according to parity, $\Pi_2^{(\text{min})}$ can always be written as
\begin{eqnarray}
	\Pi_2^{(\text{min})}(\lambda_0,\lambda_1,\lambda_2)
	&=&
	P_4(\lambda_0,\lambda_1,\lambda_2)
	+
	\sqrt{\lambda_0\lambda_1}\,Q_{01}(\lambda_0,\lambda_1,\lambda_2) \nonumber\\
	&+&
	\sqrt{\lambda_1\lambda_2}\,Q_{12}(\lambda_0,\lambda_1,\lambda_2)
	+
	\sqrt{\lambda_0\lambda_2}\,Q_{02}(\lambda_0,\lambda_1,\lambda_2),
	\label{eq:Pi2_niceform}
\end{eqnarray}
where \(P_4\) is a homogeneous polynomial of degree \(4\) in the Schmidt coefficients, and each \(Q_{ij}\) is a homogeneous polynomial of degree \(3\). Accordingly, $M_2^{(\text{max})}$ will be given by
\begin{equation}
	M_2^{(\text{max})}(\boldsymbol{\lambda})
	=
	-\ln\Bigl(
	P_4(\boldsymbol{\lambda})
	+
	\sqrt{\lambda_0\lambda_1}\,Q_{01}(\boldsymbol{\lambda})
	+
	\sqrt{\lambda_1\lambda_2}\,Q_{12}(\boldsymbol{\lambda})
	+
	\sqrt{\lambda_0\lambda_2}\,Q_{02}(\boldsymbol{\lambda})
	\Bigr).
	\label{eq:M2max_lambda}
\end{equation}

\subsection{Plotting Conventions}
\label{sec:plotting}

\begin{figure}[t]
	\centering
	\includegraphics[width=0.49\textwidth]{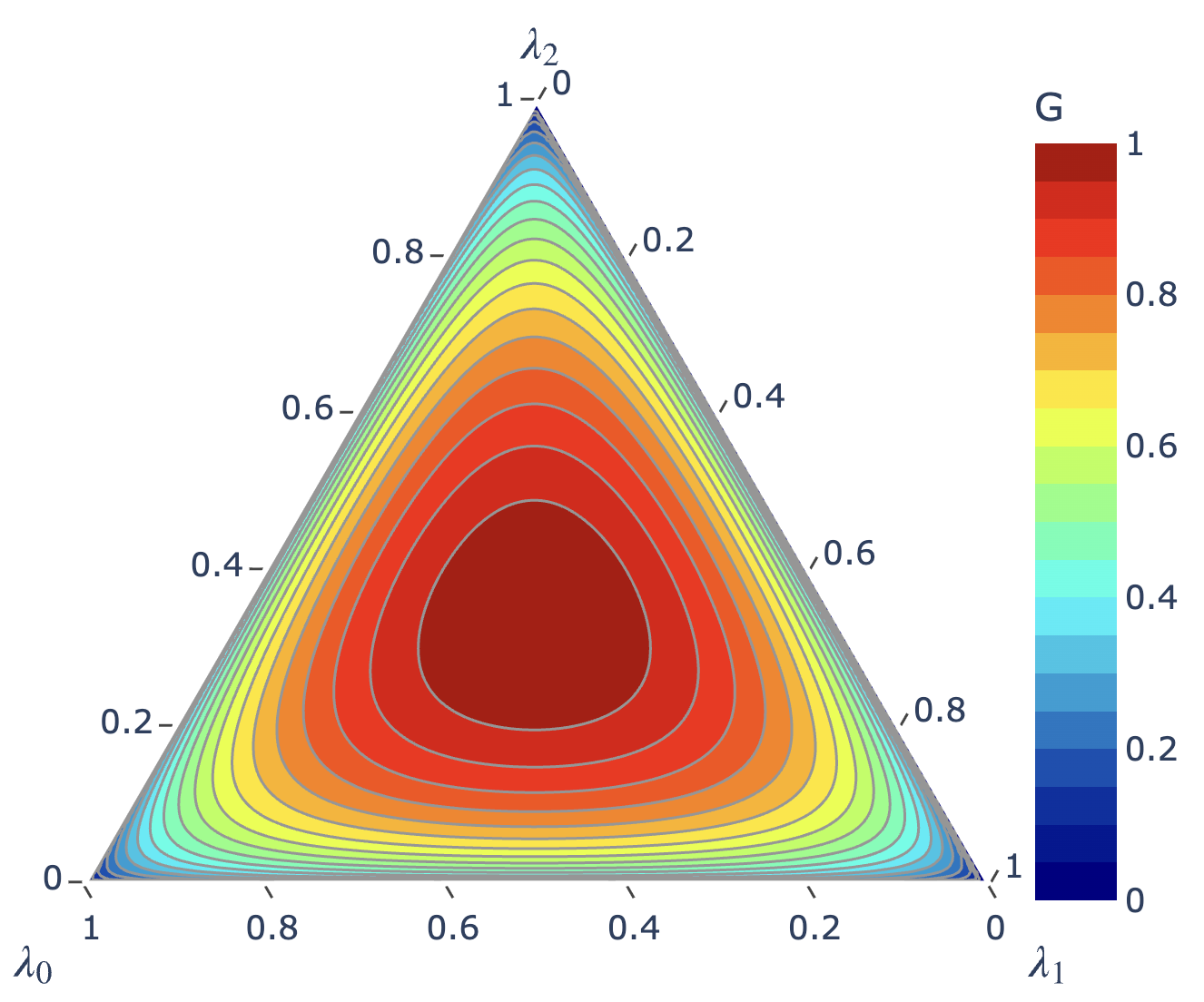}
	\includegraphics[width=0.49\textwidth]{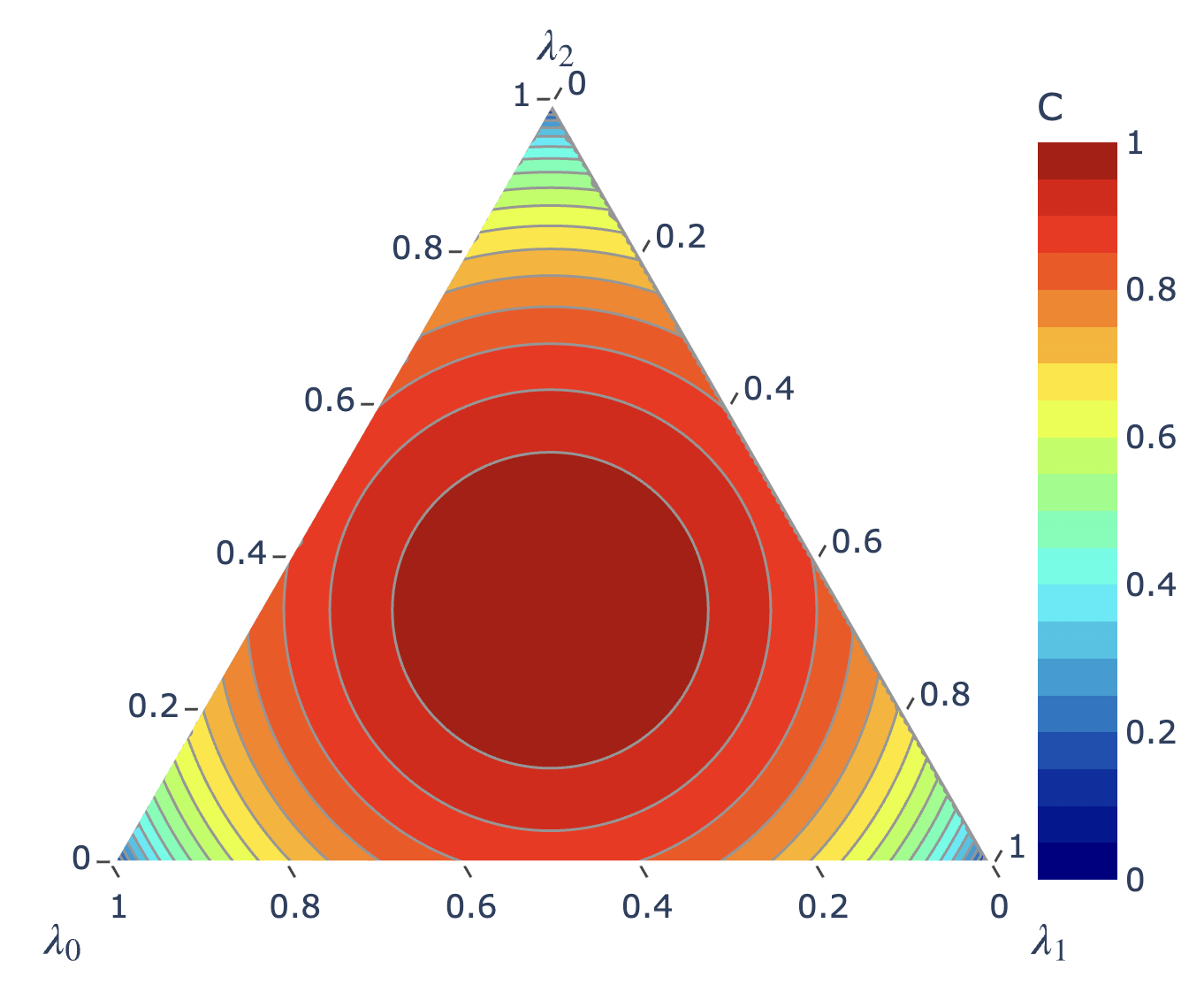}
	\caption{The G-concurrence (left) and the I-concurrence (right) as a function of $\lambda_0$, $\lambda_1$ and $\lambda_2$. }
	\label{fig:CG_xyz}
\end{figure}

Any function of the three Schmidt coefficients $\lambda_0$, $\lambda_1$ and $\lambda_2$, such as the maximal magic~(\ref{eq:M2max_lambda}), can be visualized on a ternary plot, which automatically accounts for the constraint $\lambda_0+\lambda_1+\lambda_2=1$. As an example, Figure~\ref{fig:CG_xyz} shows ternary plots of the G-concurrence (left panel) and the I-concurrence (right panel) as a function of $\lambda_0$, $\lambda_1$ and $\lambda_2$.

The ternary plots in Figure~\ref{fig:CG_xyz} nicely illustrate the different types of special two-qutrit states:
\begin{itemize}
	\item \textit{Pure product state:} If the spectrum is concentrated on only one $\lambda_i=1$ and the other two $\lambda$s vanish, then the corresponding state has Schmidt rank 1 and is completely separable into $\ket{\psi}_A \otimes \ket{\psi}_B$ and has zero entanglement. The three vertices of the equilateral triangle represent such pure product states. For example, the lower left corner corresponds to $\lambda_0=1$ and $\lambda_1=\lambda_2=0$, which is a purely separable state with both entanglement measures $C$ and $G$ equal to 0, as seen from the colorbar.
	\item \textit{Maximally entangled state:} If all Schmidt coefficients are equal, $\lambda_0 = \lambda_1 = \lambda_2 = \frac{1}{3}$, then the state has Schmidt rank 3 and is maximally entangled, with $C=1$ and $G=1$.
	\item \textit{Rank 2 state:} Each edge of a ternary plot represents 0\% of the component corresponding to the opposing vertex. For example, the lower (horizontal) edge represents states with $\lambda_2=0$ and therefore $G=0$, where the system's effective Schmidt rank drops from three to two. Traversing this edge varies the relative weights of $\lambda_0$ and $\lambda_1$, which consequently shifts the entanglement of the state as quantified by the I-concurrence $C$ (see the right panel in Figure~\ref{fig:CG_xyz}).
\end{itemize}

\begin{figure}[t]
	\centering
	\includegraphics[width=0.32\textwidth]{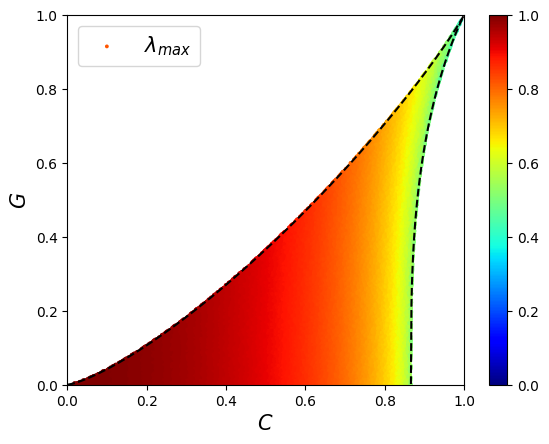}
	\includegraphics[width=0.32\textwidth]{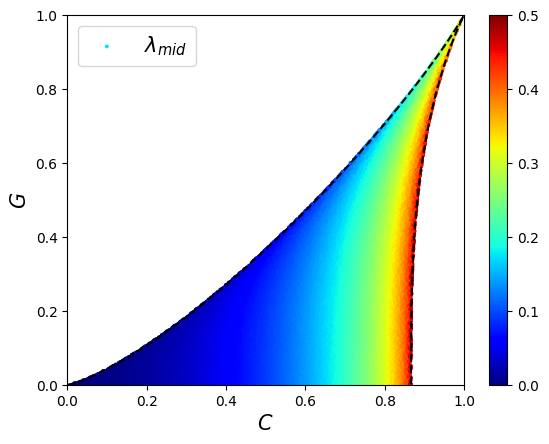}
	\includegraphics[width=0.32\textwidth]{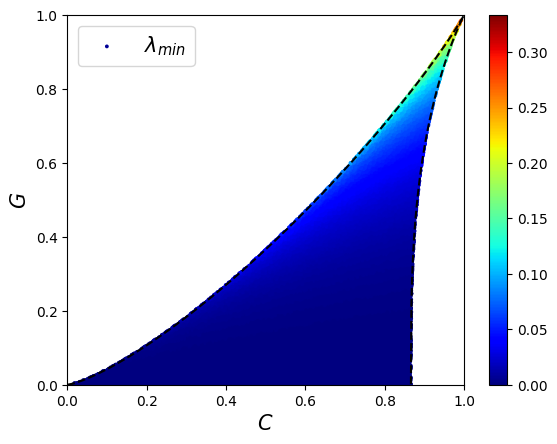}
	\caption{The Schmidt coefficients (\ref{eq:lambda_maxmidmin}) as a function of the I-concurrence $C$ and the G-concurrence $G$: $\lambda_\text{max}$ (left panel),  $\lambda_\text{mid}$ (middle panel) and   $\lambda_\text{min}$ (right panel). The dashed lines are the boundaries of the allowed region (eq.~(\ref{eq:G_range}) in the text).}
	\label{fig:lam_CG}
\end{figure}

The ternary plot representation in terms of $\lambda_0$, $\lambda_1$ and $\lambda_2$ is convenient but has two disadvantages. First, the $\lambda$ variables are not independent, but are related through the normalization condition (\ref{eq:lambda_normalization}). It would be preferable to use two independent degrees of freedom to parametrize the entanglement of the system. Second, it has a 6-fold permutation degeneracy since $M_2^{\text{max}}$ is invariant under permutations among $\lambda_0$, $\lambda_1$ and $\lambda_2$. For both of those reasons, we shall also consider an alternative representation in terms of the two {\em independent} variables $C$ and $G$:
\begin{equation}
	(\lambda_0, \lambda_1, \lambda_2)
	\leftrightarrow (C,G).
	\label{eq:lambda_to_CG}
\end{equation}
The forward transition in (\ref{eq:lambda_to_CG}) is given by eqs.~(\ref{eq:C_def}) and (\ref{eq:G_def}) and is depicted in Figure~\ref{fig:CG_xyz}, while the inverse transition is given by the solutions to the cubic equation (\ref{eq:cubic}) and is illustrated in Figure~\ref{fig:lam_CG}, where we plot the {\em ordered} Schmidt coefficients
\begin{subequations}
	\begin{eqnarray}
		\lambda_{\text{max}} &=& \max (\lambda_0, \lambda_1, \lambda_2),    \\
		\lambda_{\text{mid}} &=& \max_{\lambda_i\ne\lambda_{\text{max}}} (\lambda_0, \lambda_1, \lambda_2),    \\
		\lambda_\text{min} &=& \min (\lambda_0, \lambda_1, \lambda_2),
	\end{eqnarray}
	\label{eq:lambda_maxmidmin}
\end{subequations}
where $\lambda_{\text{max}}$, $\lambda_\text{mid}$ and $\lambda_\text{min}$ are the largest, the second largest, and the smallest Schmidt coefficients, respectively.

Figure~\ref{fig:lam_CG} reveals that the allowed $(C,G)$ parameter space is restricted and does not extend over the full unit square $(0,1)^{\otimes 2}$. One can show that the allowed region is given by
\begin{equation}
	G_\text{min}(C) \le G \le G_\text{max}(C),
	\label{eq:G_range}
\end{equation}
where
\begin{subequations}
	\begin{eqnarray}
		G_\text{min}(C) &=&
		\left\{
		\begin{array}{ll}
			0,                         & C\le \frac{\sqrt{3}}{2},  \\
			3C^2 - 2 - 2(1-C^2)^{3/2}, & C \ge \frac{\sqrt{3}}{2},
		\end{array}
		\right.\\[2mm]
		G_\text{max}(C) &=& 3C^2 - 2 + 2(1-C^2)^{3/2}.
	\end{eqnarray}
	\label{eq:Gminmax}
\end{subequations}
In Figure~\ref{fig:lam_CG}, the boundary curves~(\ref{eq:Gminmax}) are plotted with dashed lines and are seen to perfectly hug the allowed region. In what follows, we shall use interchangeably the ternary plot representation of Figure~\ref{fig:CG_xyz} and the $(C,G)$-plane representation of Figure~\ref{fig:lam_CG}.

\section{Results for Two-Qutrit States}
\subsection{The Lower Pareto Frontier: States with Minimal Magic}
\label{sec:minmagic}
To build up some intuition, in this subsection we first discuss the complementary problem of finding the {\it minimum} magic
\begin{equation}
	M_2^{(\text{min})} (\boldsymbol{\lambda})\equiv
	\min_{\boldsymbol{\varphi}_A, \boldsymbol{\varphi}_B}
	M_2(\psi(\boldsymbol{\varphi}_A, \boldsymbol{\varphi}_B, \boldsymbol{\lambda})).
	\label{eq:M2min_def}
\end{equation}
In the literature, this quantity is known as quantum nonlocal nonstabilizerness \cite{Qian:2025oit} and is typically denoted using calligraphic font and no indices, i.e., $\mathcal{M}\equiv M_2^{(\text{min})}$. However, for consistency with the rest of this paper we shall continue to denote the nonlocal nonstabilizerness with $M_2^{(\text{min})}$.

\begin{figure}[t]
	\centering
	\includegraphics[width=0.49\textwidth]{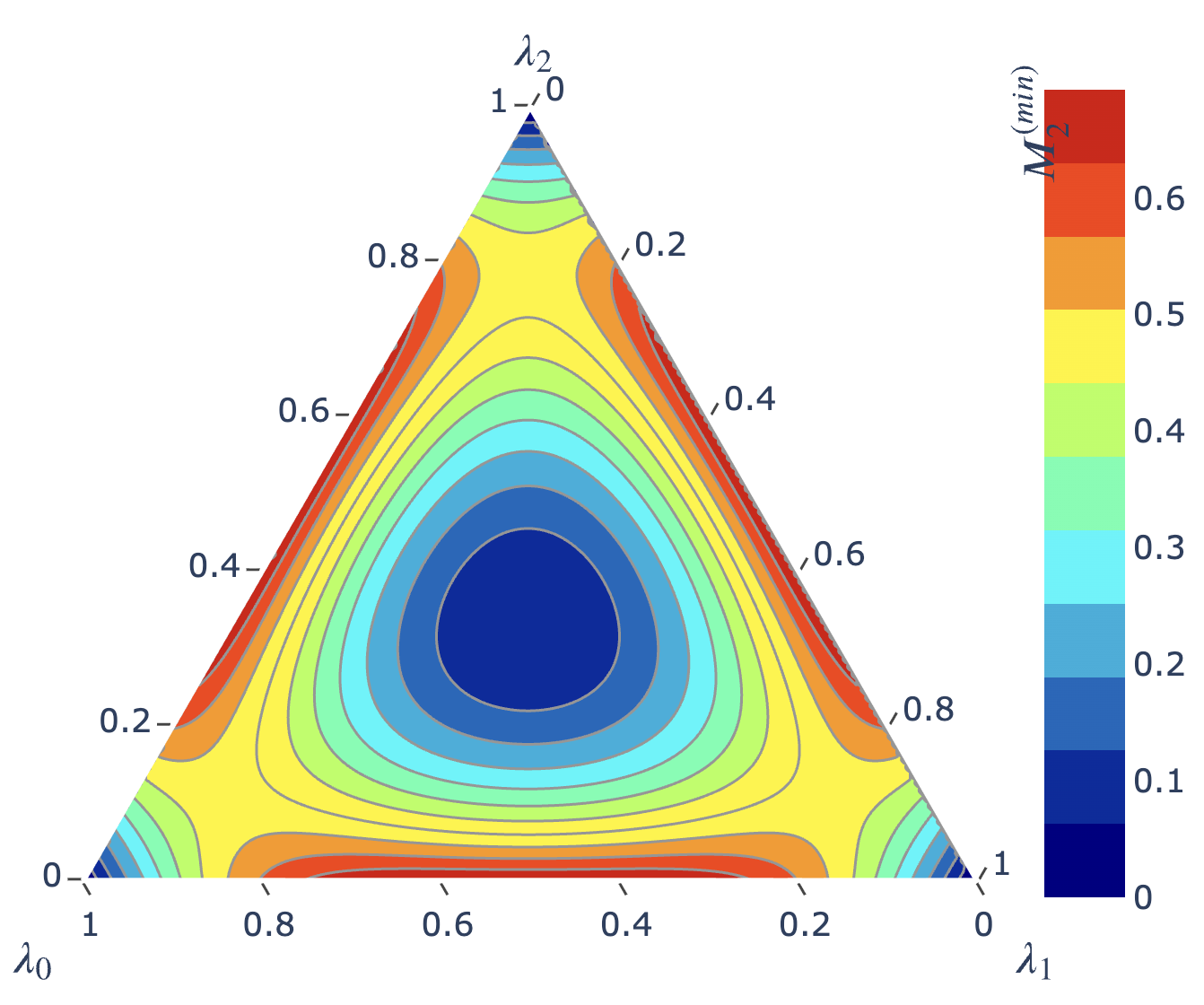}
	\includegraphics[width=0.47\textwidth]{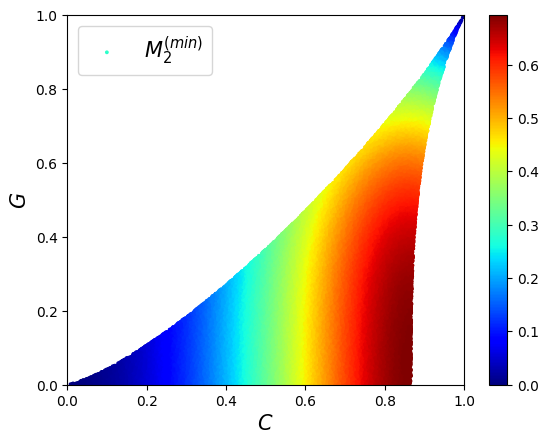}
	\caption{Plots of the minimal magic $M_2^{(\text{min})}$ defined in eq.~(\ref{eq:M2min_def}) as a function of $\boldsymbol{\lambda}$ (left panel) or as a function of the I-concurrence $C$ and the G-concurrence $G$ (right panel). }
	\label{fig:M2min}
\end{figure}

For the case of two qubits ($d=2$), it has been shown that for a given entanglement, the nonlocal nonstabilizerness is obtained for pure Schmidt states, i.e. $\vert\psi\rangle = \sqrt{\lambda_0}\vert 00\rangle + \sqrt{\lambda_1}\vert 11\rangle$ \cite{Qian:2025oit, Busoni:2026lvp}. Analytical expressions for the nonlocal nonstabilizerness as a function of entanglement are also available \cite{Qian:2025oit,Busoni:2026lvp,Roman:2026mcy}. Ref.~\cite{Busoni:2026lvp} conjectured that pure Schmidt states continue to define the nonlocal nonstabilizerness for any prime $d$, and showed numerically that the conjecture holds for the case of qutrits ($d=3$) and ququints ($d=5$). Using the Schmidt ansatz, Ref.~\cite{Busoni:2026lvp} also derived an analytical formula for $M_2^{(\text{min})}$ as a function of the Schmidt coefficients (see equation~(43) in~\cite{Busoni:2026lvp}), which in our notation can be rewritten as
\begin{align}
	\nonumber
	M_2^{(\text{min})}(\lambda_0,\lambda_1,\lambda_2) =
	 & -\ln \Biggl[
	1 -2\sum_{i=0}^2 \lambda_i^2 +2\left(\sum_{i=0}^2 \lambda_i^2\right)^2                                                                                     \\
	 & + 4\lambda_0\lambda_1\lambda_2 \left( 1+2\bigl( \sqrt{\lambda_0\lambda_1} +\sqrt{\lambda_1\lambda_2} +\sqrt{\lambda_2\lambda_0} \bigr) \right) \Biggr].
	\label{eq:M2min_lambdas}
\end{align}
This function is visualized in the left panel of Figure~\ref{fig:M2min} as a ternary plot (see also the analogous Figure 1 in \cite{Busoni:2026lvp}). It can be seen that the maximal value of $M_2^{(\text{min})}$ is obtained for rank-two spectra, $(\lambda_0,\lambda_1,\lambda_2)=(0.5,0.5,0)$ and permutations \cite{Busoni:2026lvp}.

As discussed in Section~\ref{sec:plotting}, we find it instructive to express the magic in terms of two (independent) entanglement degrees of freedom instead of the three (constrained) Schmidt coefficients $\lambda_0$, $\lambda_1$, and $\lambda_2$. In this case, it is convenient to use the I-concurrence $C$ from eq.~(\ref{eq:C_def}) and the negativity $N$ from eq.~(\ref{eq:N_def}) as independent variables, which allows us to rewrite eq.~(\ref{eq:M2min_lambdas}) as
\begin{equation}
	M_2^{(\text{min})}(C,N)=
	- \ln \left[
		\frac{1}{2} + \frac{1}{2} \left(1-\frac{4}{3}C^2\right)^2
		+\left(N^2-\frac{1}{3}C^2\right)^2\right].
	\label{eq:M2min_CG}
\end{equation}
This result makes it clear that a) the maximal value of $M_2^{(\text{min})}$ is obtained at $C=\frac{\sqrt{3}}{2}\approx 0.866$ and $N=\frac{C}{\sqrt{3}}=\frac{1}{2}$, which causes the last two positive-definite terms under the logarithm to vanish, and b) that the resulting maximal value of $M_2^{(\text{min})}$ is equal to $\ln2$:
\begin{equation}
	\max_{\boldsymbol{\lambda}}\left(M_2^{(\text{min})}(\lambda_0,\lambda_1,\lambda_2)\right)  =\ln 2.
\end{equation}

The right panel in Figure~\ref{fig:M2min} shows the $(C,G)$ representation of the nonlocal nonstabilizerness function $M_2^{(\text{min})}$. This plot can be compared to the $(\lambda_0,\lambda_1,\lambda_2)$ ternary plot shown in the left panel. The minimum at $C=G=0$ corresponds to the three minima at the corners of the ternary plot, which represent product states with no entanglement. The minimum at $C=G=1$ maps to the maximally entangled point in the center of the ternary plot, $(\lambda_0,\lambda_1,\lambda_2)=\left(\frac{1}{3},\frac{1}{3},\frac{1}{3}\right)$. As already discussed above, the maximum of $M_2^{(\text{min})}$ is found at $N=\frac{C}{\sqrt{3}}$, which implies that $G=0$ in light of eq.~(\ref{eq:NCG}). The plot in the right panel of Figure~\ref{fig:M2min} reveals that the maximum indeed appears at this special vertex point with maximal I-concurrence and zero G-concurrence.

\subsection{The Upper Pareto Frontier: States with Maximal Magic}
\label{sec:maxmagic}

\begin{figure}[t]
	\centering
	\includegraphics[trim={0cm 0cm 0cm 0cm}, clip, width=0.55\linewidth]{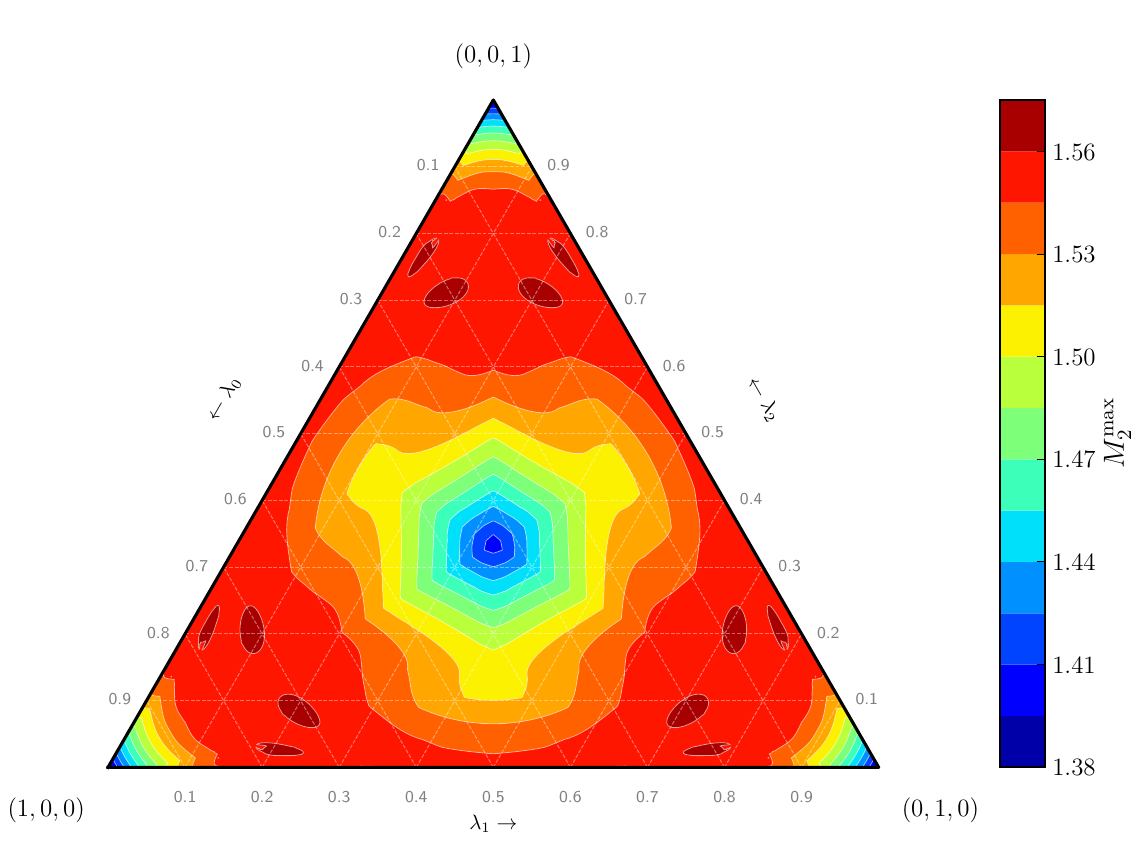}
	\hfill
	\includegraphics[width=0.43\linewidth]{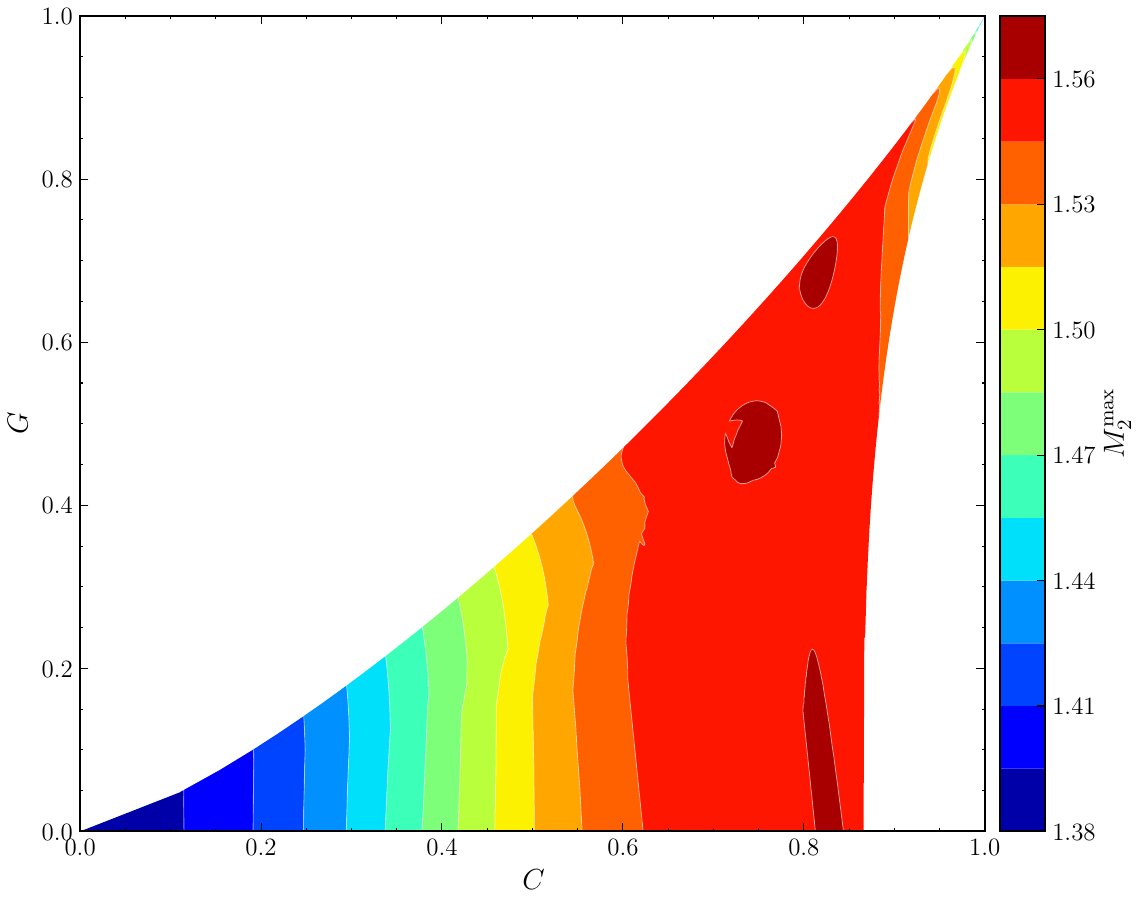}
	\caption{The same as Figure~\ref{fig:M2min}, but for the numerically optimized maximal magic $M_2^{\text{max}}$ as a function of $\boldsymbol{\lambda}$ (left panel) or $C$ and $G$ (right panel).
	}
	\label{fig:M2max}
\end{figure}

We now turn to our main task, the analytical exploration of the maximal magic $M_2^{(\text{max})}$ defined in eq.~(\ref{eq:M2max_def}). In order to gain some initial understanding of the $M_2^{(\text{max})}$ landscape, we perform the optimization in eq.~(\ref{eq:M2max_def}) numerically by starting from the Schmidt decomposition~(\ref{eq:Schmidt_decomposition_LU}) and maximizing the magic over the parameters $\boldsymbol{\varphi}_A$ and $\boldsymbol{\varphi}_B$ of the local unitaries, i.e. we compute\footnote{The optimization in eq.~(\ref{eq:M2maximization}) is straightforward, but does have some subtleties which are discussed in Appendix~\ref{app:numerics}, together with our validation cross-checks.} the quantity
\begin{equation}
	\max_{\boldsymbol{\varphi}_A, \boldsymbol{\varphi}_B}
	M_2(
	\left[U_A(\boldsymbol{\varphi}_A) \otimes U_B\left(\boldsymbol{\varphi}_B)\right]\ket{\psi_{\mathrm{S}}(\boldsymbol{\lambda})}\right).
	\label{eq:M2maximization}
\end{equation}
The results are plotted in  Figure~\ref{fig:M2max} in complete analogy to Figure~\ref{fig:M2min}.

Figure~\ref{fig:M2max} reveals that the magic is maximized at three different locations in the $(C,G)$ plane which correspond to $3\times 3!=18$ different locations in the $\boldsymbol{\lambda}$ ternary plot. We find that 6 of the 18 maxima on the ternary plot happen to be exactly on the edges of the plot, where one of the $\lambda$'s is equal to 0, while the other two $\lambda$'s are in a ratio of $2:1$. Given the plot's resolution, these 6 maxima are difficult to see on the ternary plot in the left panel, but their existence is clearly evident in the $(C,G)$ plot in the right panel near $G=0$ and $C=\sqrt{2/3}\approx 0.8165$.

\begin{table}[t]
	\caption{\label{tab:parameters}
	Parameters describing the locations of the three maxima of the maximal magic $M_2^{(\text{max})}(\boldsymbol{\lambda})$. Note that the values for all listed variables are known analytically, but if the corresponding formula is too lengthy to fit in the table, we just list the numerical value.}
	\begin{ruledtabular}
		\begin{tabular}{cccc}
			Variable             & Maximum 1                                             & Maximum 2                                                                                                    & Maximum 3                       \\
			\colrule
			$\lambda_\text{max}$ & 0.71238601...                                         & $\frac{2}{3}$                                                                                                & 0.76494263...                   \\
			$\lambda_\text{mid}$ & 0.20168972...                                         & $\frac{1}{3}$                                                                                                & 0.20936118...                   \\
			$\lambda_\text{min}$ & 0.08592427...                                         & $0$                                                                                                          & 0.02569619...                   \\[1mm]
			\hline
			$C$                  & $\frac{\sqrt{2}}{\sqrt{3}}$
			                     & $\frac{\sqrt{2}}{\sqrt{3}}$                           & $\frac{\sqrt{5}}{3}$                                                                                                                           \\[1mm]
			$G$                  & $3^{-\frac{1}{3}}$                                    & 0                                                                                                            & $9^{-\frac{1}{3}}$              \\
			$N$                  & $\frac{4}{3\sqrt{3}}\cos \left(\frac{\pi}{18}\right)$
			                     & $\frac{\sqrt{2}}{3}$                                  & $\frac{1}{9}\left(1+4\sqrt{2}\cos\left(\frac{1}{3} \arccos\left(\frac{-1}{2\sqrt{2}}\right) \right)\right) $
			\\ [2mm]
			\colrule
			$N_{-++}$            & $0$                                                   & $-\frac{\sqrt{2}}{3}$                                                                                        & $-0.18663931...$                \\
			$N_{+-+}$            & $0.49481813...$                                       & $\frac{\sqrt{2}}{3}$                                                                                         & $0.46703981...$                 \\
			$N_{++-}$            & $0.26328723...$                                       & $\frac{\sqrt{2}}{3}$                                                                                         & $\frac{1}{3}$                   \\
			\colrule
			$M_2^{(\text{max})}$ & $\ln\left(\frac{81}{17}\right)$                       & $\ln\left(\frac{81}{17}\right)$                                                                              & $\ln\left(\frac{81}{17}\right)$ \\
		\end{tabular}
	\end{ruledtabular}
\end{table}

In what follows, we shall focus on the three maxima in the $(C,G)$ plane and describe each one in detail: we shall specify the states $\psi$ which correspond to each maximum and derive the analytical expression for $M_2^{(\text{max})}(\boldsymbol{\lambda})$ which is valid in the vicinity of that maximum. The states $\psi$ will be specified via the local unitaries $U$ and $V^\dagger$ entering the singular value decomposition of the $3\times 3$ complex matrix $A_{ij}\equiv a_{ij}$ of the amplitudes $a_{ij}$ appearing in eq.~(\ref{eq:generic_psi_def}):
\begin{equation}
	A = U \, \Sigma \, V^\dagger,
	\label{eq:svd}
\end{equation}
where $\Sigma \equiv \text{diag}(\sqrt{\lambda_\text{max}}, \sqrt{\lambda_\text{mid}}, \sqrt{\lambda_\text{min}})$ is a diagonal $3\times 3$ matrix constructed out of the ordered Schmidt coefficients $\lambda_\text{max}$, $\lambda_\text{mid}$ and $\lambda_\text{min}$.

For reference, the values of the relevant parameters for each of the three maxima are collected in Table~\ref{tab:parameters} and will be derived and discussed in the remainder of this section. Table~\ref{tab:parameters} also lists the values for the three helper functions
\begin{subequations}
	\begin{eqnarray}
		N_{-++} &=&
		-\sqrt{\lambda_\text{max}}\sqrt{\lambda_\text{mid}}
		+\sqrt{\lambda_\text{mid}}\sqrt{\lambda_\text{min}}
		+\sqrt{\lambda_\text{max}}\sqrt{\lambda_\text{min}},\\
		N_{+-+} &=&
		+\sqrt{\lambda_\text{max}}\sqrt{\lambda_\text{mid}}
		-\sqrt{\lambda_\text{mid}}\sqrt{\lambda_\text{min}}
		+\sqrt{\lambda_\text{max}}\sqrt{\lambda_\text{min}},\\
		N_{++-} &=&
		+\sqrt{\lambda_\text{max}}\sqrt{\lambda_\text{mid}}
		+\sqrt{\lambda_\text{mid}}\sqrt{\lambda_\text{min}}
		-\sqrt{\lambda_\text{max}}\sqrt{\lambda_\text{min}},
	\end{eqnarray}
	\label{eq:f_def}
\end{subequations}
which are related to the negativity in eq.~(\ref{eq:N_def}) as
\begin{equation}
	N_{-++} + N_{+-+} + N_{++-}= N.
\end{equation}

\subsubsection{The first maximum: \texorpdfstring{$G^3=1/3$}{G3=1/3}}

The states which give the first maximum have a very specific structure: their local unitaries $U$ and $V$ depend on the Schmidt coefficients at the maximum themselves. For concreteness, let us start with the specific permutation
$\lambda_0^{(1)}=\lambda_{\text{max}}^{(1)}$,
$\lambda_1^{(1)}=\lambda_{\text{mid}}^{(1)}$, and
$\lambda_2^{(1)}=\lambda_{\text{min}}^{(1)}$.
As an illustration, one such state is given by
\begin{eqnarray}
	U (\boldsymbol{\lambda}^{(1)})&=&
	\begin{pmatrix}
		-\sqrt{\lambda_0^{(1)}}                  &
		\sqrt{\lambda_2^{(1)}}e^{-120^\circ\, i} &
		\sqrt{\lambda_1^{(1)}}e^{120^\circ\, i}    \\
		\sqrt{\lambda_2^{(1)}}e^{120^\circ\, i}  &
		-\sqrt{\lambda_1^{(1)}}                  &
		\sqrt{\lambda_0^{(1)}}e^{-120^\circ\, i}   \\
		\sqrt{\lambda_1^{(1)}}e^{60^\circ\, i}   &
		\sqrt{\lambda_0^{(1)}}e^{-60^\circ\, i}  &
		-\sqrt{\lambda_2^{(1)}}                    \\
	\end{pmatrix}, \label{eq:U1def}\\[3mm]
	V^\dagger (\boldsymbol{\lambda}^{(1)}) &=&
	\begin{pmatrix}
		-\sqrt{\lambda_2^{(1)}}                 &
		\sqrt{\lambda_1^{(1)}}e^{120^\circ\, i} &
		\sqrt{\lambda_0^{(1)}}e^{120^\circ\, i}   \\
		\sqrt{\lambda_0^{(1)}}                  &
		\sqrt{\lambda_2^{(1)}}e^{-60^\circ\, i} &
		\sqrt{\lambda_1^{(1)}}e^{120^\circ\, i}   \\
		\sqrt{\lambda_1^{(1)}}                  &
		\sqrt{\lambda_0^{(1)}}e^{120^\circ\, i} &
		\sqrt{\lambda_2^{(1)}}e^{-60^\circ\, i}   \\
	\end{pmatrix}, \label{eq:V1def}
\end{eqnarray}
where $\lambda_0^{(1)}= 0.71238601...$,
$\lambda_1^{(1)}= 0.20168972...$ and
$\lambda_2^{(1)}= 0.08592427...$
are the Schmidt coefficients which are listed in the first column of Table~\ref{tab:parameters}. Their numerical values can be obtained {\em exactly} as the solutions to the cubic equation (\ref{eq:cubic}) with $C=\sqrt{2}/\sqrt{3}$ and $G=3^{-1/3}$, as shown in the first column of Table~\ref{tab:parameters}. Note that the complex phases in eqs.~(\ref{eq:U1def}) and (\ref{eq:V1def}) are written in degrees and are multiples of $60^\circ$. The overall phases have been chosen so that the diagonal of $U$ and the first column of $V^\dagger$ are real.

In order to find the maximal $M_2$ away from the maximum, we form the state $\psi$ via eq.~(\ref{eq:svd}):
\begin{equation}
	A = U \left(\boldsymbol{\lambda}^{(1)}\right)\, \Sigma\left(\boldsymbol{\lambda}\vphantom{^1}\right) \, V^\dagger \left(\boldsymbol{\lambda}^{(1)}\right),
	\label{eq:A1state}
\end{equation}
where we have frozen the local unitaries to their values at the maximum, so that the $\boldsymbol{\lambda}$ dependence enters only through the diagonal matrix $\Sigma$. Substituting the result into the definition of $\Pi_2$ from eq.~(\ref{eq:purity}), we find that the resulting polynomial (\ref{eq:Pi2_polynomial}) in $\boldsymbol{\lambda}$ is given by
\begin{eqnarray}
	{\cal P}^{(1)} (\lambda_0,\lambda_1,\lambda_2) &=&
	\frac{17}{81} + \frac{1}{81}\Bigl[8\left(\lambda_0^4+\lambda_1^4+\lambda_2^4\right)
		+ 96\left(\lambda_0^2\lambda_1^2+\lambda_1^2\lambda_2^2+\lambda_0^2\lambda_2^2\right) \nonumber\\
		&-&28 \left(\lambda_0\lambda_1^3+\lambda_1\lambda_0^3
		+\lambda_1\lambda_2^3+\lambda_2\lambda_1^3
		+\lambda_0\lambda_2^3+\lambda_2\lambda_0^3\right)
		+204\, \lambda_0\lambda_1\lambda_2 \nonumber\\
		&-&368\left(
		\lambda_0\lambda_1^{\frac{3}{2}}\lambda_2^{\frac{3}{2}}
		+\lambda_0^{\frac{3}{2}}\lambda_1\lambda_2^{\frac{3}{2}}
		-\lambda_0^{\frac{3}{2}}\lambda_1^{\frac{3}{2}}\lambda_2
		\right) \nonumber \\
		&-& 24\left(
		\lambda_0^3\lambda_1^{\frac{1}{2}}\lambda_2^{\frac{1}{2}}
		+\lambda_0^{\frac{1}{2}}\lambda_1^3\lambda_2^{\frac{1}{2}}
		-\lambda_0^{\frac{1}{2}}\lambda_1^{\frac{1}{2}}\lambda_2^3
		\right) \nonumber \\
		&-& 96\left(
		\lambda_0^2\lambda_1^{\frac{1}{2}}\lambda_2^{\frac{3}{2}}
		+\lambda_0^{\frac{1}{2}}\lambda_1^2\lambda_2^{\frac{3}{2}}
		-\lambda_0^{\frac{1}{2}}\lambda_1^{\frac{3}{2}}\lambda_2^2
		+\lambda_0^2\lambda_1^{\frac{3}{2}}\lambda_2^{\frac{1}{2}}
		+\lambda_0^{\frac{3}{2}}\lambda_1^2\lambda_2^{\frac{1}{2}}
		-\lambda_0^{\frac{3}{2}}\lambda_1^{\frac{1}{2}}\lambda_2^2
		\right) \nonumber \\
		&+& 24 \left(
		\lambda_0\lambda_1^{\frac{1}{2}}\lambda_2^{\frac{5}{2}}
		+\lambda_0\lambda_1^{\frac{5}{2}}\lambda_2^{\frac{1}{2}}
		+\lambda_0^{\frac{1}{2}}\lambda_1\lambda_2^{\frac{5}{2}}
		+\lambda_0^{\frac{5}{2}}\lambda_1\lambda_2^{\frac{1}{2}}
		-\lambda_0^{\frac{1}{2}}\lambda_1^{\frac{5}{2}}\lambda_2
		-\lambda_0^{\frac{5}{2}}\lambda_1^{\frac{1}{2}}\lambda_2
		\right)\nonumber \\
		&+& 16 \left(
		\lambda_0^{\frac{3}{2}}\lambda_2^{\frac{5}{2}}
		+\lambda_0^{\frac{5}{2}}\lambda_2^{\frac{3}{2}}
		+\lambda_1^{\frac{3}{2}}\lambda_2^{\frac{5}{2}}
		+\lambda_1^{\frac{5}{2}}\lambda_2^{\frac{3}{2}}
		-\lambda_0^{\frac{3}{2}}\lambda_1^{\frac{5}{2}}
		-\lambda_0^{\frac{5}{2}}\lambda_1^{\frac{3}{2}}
		\right)
		\Bigr],
	\label{eq:Pi2_case1}
\end{eqnarray}
and that the maximal magic in the vicinity of the maximum is given by the function
\begin{equation}
	f^{(1)}(\lambda_0,\lambda_1,\lambda_2) = - \ln {\cal P}^{(1)} (\lambda_0,\lambda_1,\lambda_2).
\end{equation}
In (\ref{eq:Pi2_case1}) we have taken advantage of the normalization condition (\ref{eq:lambda_normalization}) to isolate a numerical constant $\frac{17}{81}$ in the first line.
It can be shown that a) the remaining $\lambda$-dependent expression within the square parentheses in (\ref{eq:Pi2_case1}) is non-negative, b) its minimum is zero, and c) that minimum is achieved at $\boldsymbol{\lambda}=\boldsymbol{\lambda}^{(1)}$. Therefore, the minimum value of ${\cal P}^{(1)}$ defined in eq.~(\ref{eq:Pi2_case1}) is ${\cal P}^{(1)}(\lambda_0^{(1)},\lambda_1^{(1)},\lambda_2^{(1)})=\frac{17}{81}$. Correspondingly, as shown in Table~\ref{tab:parameters}, the largest value of the maximal magic $f^{(1)}=-\ln {\cal P}^{(1)}$ is $\ln\frac{81}{17}\approx 1.561$ and is obtained also at $\boldsymbol{\lambda}=\boldsymbol{\lambda}^{(1)}$. Our result improves on the previous bound of $\ln 5\approx 1.609$ given by (\ref{eq:Cuffaro_bound}) and on the numerically derived result of $\ln (2^{2.23379})\approx 1.548$ quoted in Ref.~\cite{Chernyshev:2024pqy}.

\begin{figure}[t]
	\centering
	\includegraphics[trim={0cm 0.0cm 0cm 0.0cm}, clip, width=0.32\linewidth]{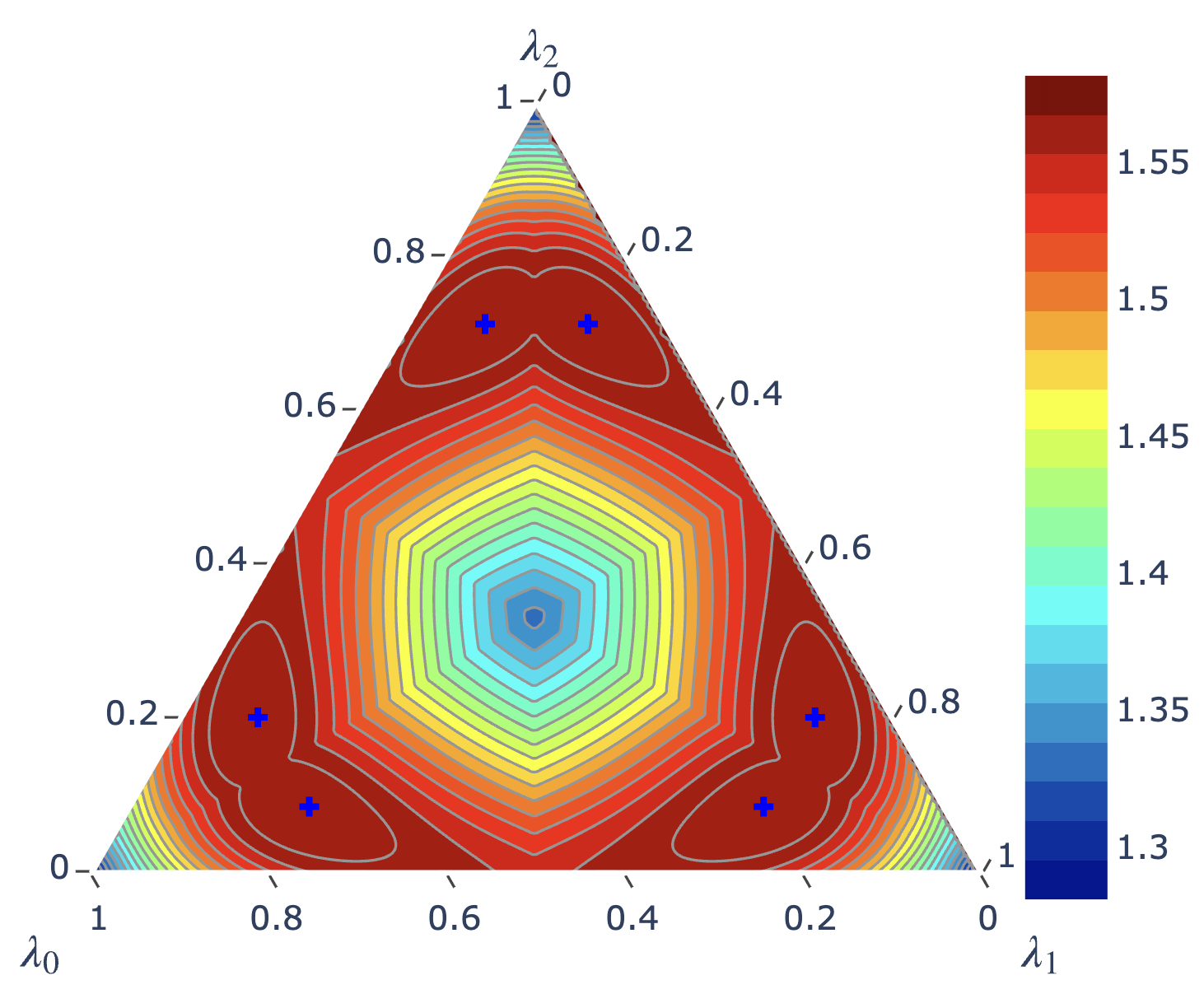}
	\includegraphics[trim={0cm 0.0cm 0cm 0.0cm}, clip, width=0.32\linewidth]{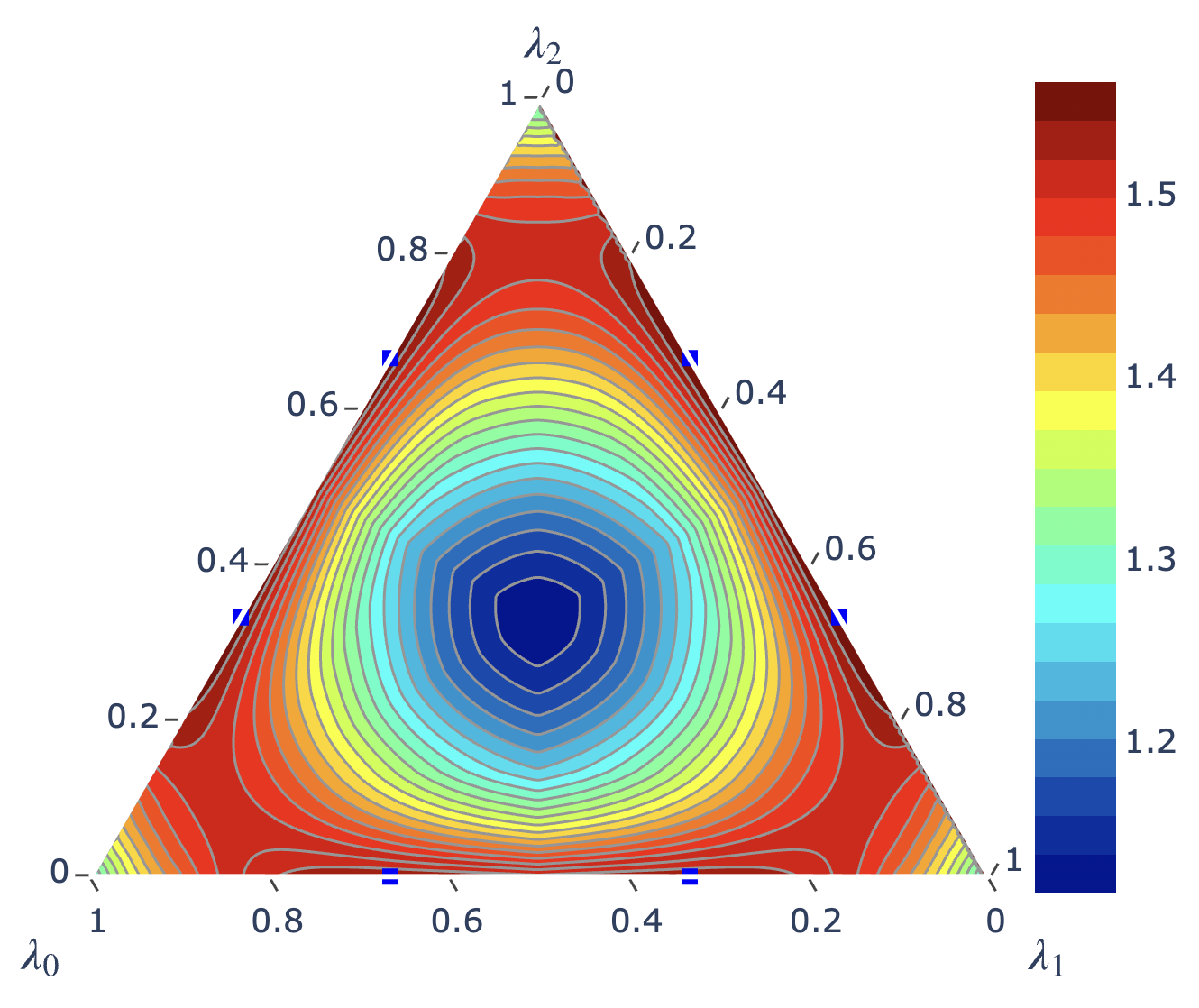}        \includegraphics[trim={0cm 0.0cm 0cm 0.0cm}, clip, width=0.32\linewidth]{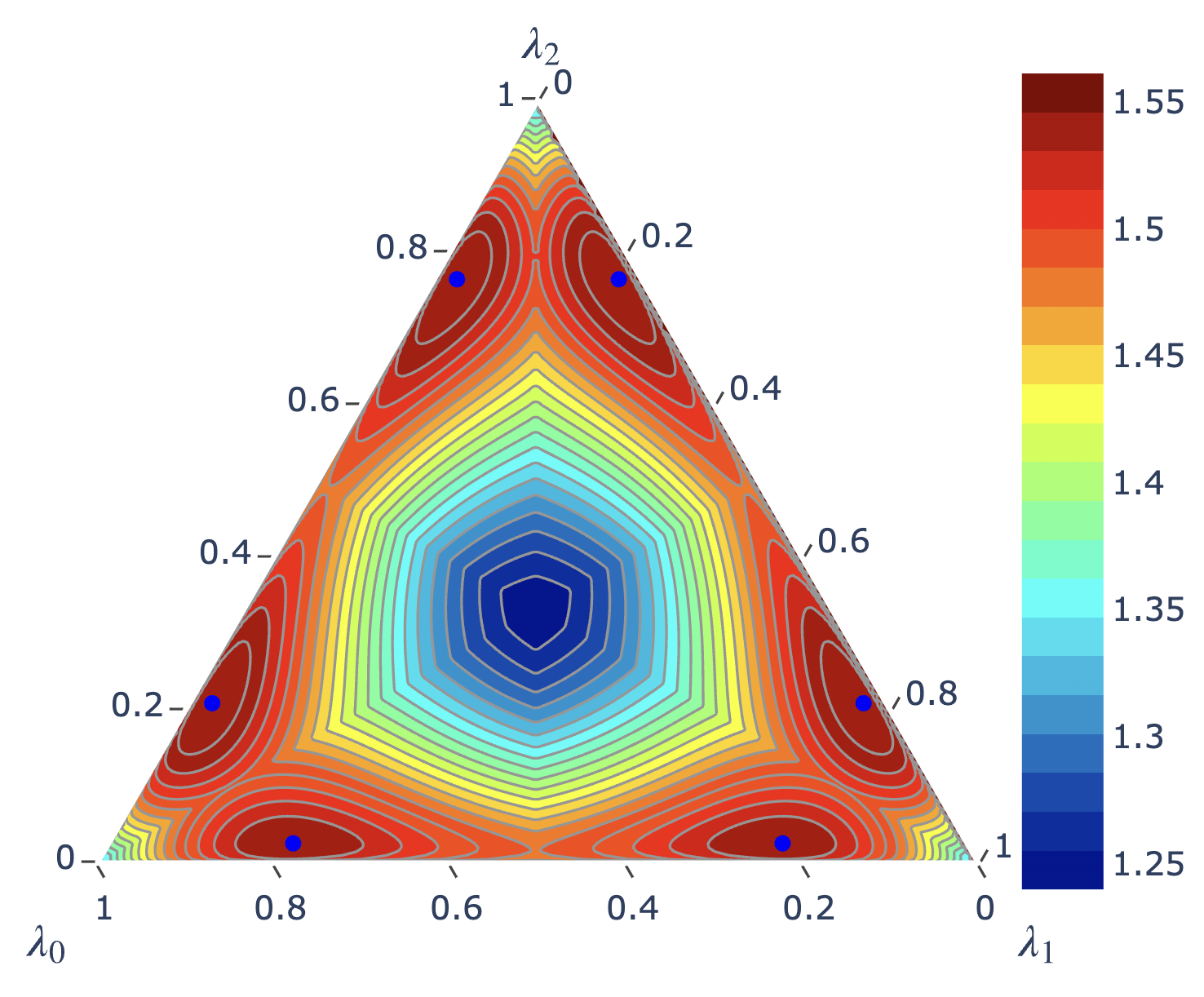}
	\caption{The maximal magic $M_2^{(i)}$, $i=1,2,3$, given by the corresponding analytical expression (\ref{eq:Mmax_case1}), (\ref{eq:Mmax_case2}) and (\ref{eq:Mmax_case3}), as a function of $\boldsymbol{\lambda}$. The symbols mark the locations of the 6 maxima in each case.
	}
	\label{fig:fmaxfunctions}
\end{figure}

The polynomial ${\cal P}^{(1)} (\lambda_0,\lambda_1,\lambda_2)$ is symmetric with respect to interchanging $\lambda_0 \leftrightarrow \lambda_1$, therefore a second degenerate maximum of $f^{(1)}$ exists at $\boldsymbol{\lambda} = (\lambda_1^{(1)},\lambda_0^{(1)},\lambda_2^{(1)})$. In order to obtain the other 4 maxima, we need to consider $f^{(1)}(\lambda_2,\lambda_1,\lambda_0)$ and
$f^{(1)}(\lambda_0,\lambda_2,\lambda_1)$. Then all 6 maxima can be described by the function
\begin{equation}
	M_2^{(1)}(\lambda_0,\lambda_1,\lambda_2) =
	\max \left(
	f^{(1)}(\lambda_0,\lambda_1,\lambda_2),
	f^{(1)}(\lambda_2,\lambda_1,\lambda_0),
	f^{(1)}(\lambda_0,\lambda_2,\lambda_1)
	\right),
	\label{eq:Mmax_case1}
\end{equation}
which is plotted in the left panel of Figure~\ref{fig:fmaxfunctions}.

\subsubsection{The second maximum: \texorpdfstring{$G=0$}{G=0}}

The states which give the second maximum shown in Table~\ref{tab:parameters} have a similar structure. One specific state at the maximum is
\begin{eqnarray}
	U (\boldsymbol{\lambda}^{(1)})&=&
	\begin{pmatrix}
		\sqrt{\lambda_0^{(1)}}e^{-120^\circ\, i} &
		\sqrt{\lambda_2^{(1)}}e^{60^\circ\, i}   &
		\sqrt{\lambda_1^{(1)}}e^{-120^\circ\, i}   \\
		\sqrt{\lambda_2^{(1)}}e^{-60^\circ\, i}  &
		\sqrt{\lambda_1^{(1)}}e^{120^\circ\, i}  &
		\sqrt{\lambda_0^{(1)}}e^{120^\circ\, i}    \\
		\sqrt{\lambda_1^{(1)}}                   &
		\sqrt{\lambda_0^{(1)}}                   &
		-\sqrt{\lambda_2^{(1)}}                    \\
	\end{pmatrix}
	,
	\label{eq:U3def}\\[2mm]
	V^\dagger &=&
	\begin{pmatrix}
		\frac{-1}{\sqrt{6}}                     &
		\frac{1}{\sqrt{6}}\, e^{60^\circ\, i}   &
		\frac{\sqrt{2}}{\sqrt{3}}                 \\
		\frac{1}{\sqrt{3}}                      &
		\frac{1}{\sqrt{3}}\, e^{-120^\circ\, i} &
		\frac{1}{\sqrt{3}}                        \\
		\frac{1}{\sqrt{2}}                      &
		\frac{1}{\sqrt{2}}\, e^{60^\circ\, i}   &
		0                                         \\
	\end{pmatrix}. \label{eq:V3def}
\end{eqnarray}
Note that $U$ still depends on the values of the Schmidt coefficients $\left(\lambda_0^{(1)}, \lambda_1^{(1)}, \lambda_2^{(1)}\right)$ {\it at the first maximum}, hence the upper index ``(1)'' on $\lambda_i$. On the other hand, $V$ is comprised of numerical constants only. The corresponding purity polynomial is given by
\begin{eqnarray}
	{\cal P}^{(2)} (\boldsymbol{\lambda}) &=&
	\frac{1}{9}\,\biggl[\,
		\frac{5}{2}\lambda_0^4
		+ 5\lambda_1^4
		+ \frac{5}{2}\lambda_2^4
		+ 5\lambda_0^3\lambda_1
		+ 4\lambda_0^3\lambda_2
		+ 7\lambda_0\lambda_2^3
		+ 4\lambda_1\lambda_2^3
		\nonumber \\
		&+& 21\lambda_0^2\lambda_1^2
		+ 21\lambda_0^2\lambda_2^2
		+ 21\lambda_1^2\lambda_2^2
		+ 24\lambda_0^2\lambda_1\lambda_2
		+ 60\lambda_0\lambda_1^2\lambda_2
		+ 42\lambda_0\lambda_1\lambda_2^2 \nonumber \\
		&+& 2\sqrt{2}\,\sqrt{\lambda_0\lambda_1}\,\lambda_0\lambda_1\big(8\lambda_2-\lambda_0\big)
		\biggr].
	\label{eq:Pi2_case3}
\end{eqnarray}
This polynomial is minimized at $\boldsymbol{\lambda}=\boldsymbol{\lambda}^{(2)}=(\frac{2}{3},\frac{1}{3},0)$, the values shown in the second column of Table~\ref{tab:parameters}. The corresponding minimum of ${\cal P}^{(2)}$ at that location is equal to $\frac{17}{81}$.

The maximal magic in the vicinity of any of the six maxima of type ``(2)'' is then given by
\begin{equation}
	M_2^{(2)}(\lambda_0,\lambda_1,\lambda_2) =
	\max_{\substack{\text{all permutations} \\ \text{of } i, j, k}}
	\left(
	f^{(2)}(\lambda_i,\lambda_j,\lambda_k)
	\right),
	\label{eq:Mmax_case2}
\end{equation}
where this time the helper function is
\begin{equation}
	f^{(2)}(\lambda_0,\lambda_1,\lambda_2) = - \ln {\cal P}^{(2)} (\lambda_0,\lambda_1,\lambda_2).
\end{equation}
The value of the magic at the maximum is once again equal to $\ln\frac{81}{17}$, indicating that this maximum is degenerate with the one considered previously at $\boldsymbol{\lambda}=\boldsymbol{\lambda}^{(1)}$. The function (\ref{eq:Mmax_case2}) is plotted in the middle panel of Figure~\ref{fig:fmaxfunctions}.

\subsubsection{The third maximum: \texorpdfstring{$G^3=1/9$}{G3=1/9}}

In the resource theory of magic, Clifford unitaries constitute the set of free operations. Consequently, any well-defined measure of magic (or non-stabilizerness) must be a monotone that is invariant under Clifford transformations. Having found one maximum of $M_2$, we can easily obtain the others by acting with Clifford operators. For example, we can get to the third maximum by applying the SUM gate operator
\begin{equation}
	\begin{pmatrix}
		1 & 0 & 0 & 0 & 0 & 0 & 0 & 0 & 0 \\
		0 & 1 & 0 & 0 & 0 & 0 & 0 & 0 & 0 \\
		0 & 0 & 1 & 0 & 0 & 0 & 0 & 0 & 0 \\
		0 & 0 & 0 & 0 & 0 & 1 & 0 & 0 & 0 \\
		0 & 0 & 0 & 1 & 0 & 0 & 0 & 0 & 0 \\
		0 & 0 & 0 & 0 & 1 & 0 & 0 & 0 & 0 \\
		0 & 0 & 0 & 0 & 0 & 0 & 0 & 1 & 0 \\
		0 & 0 & 0 & 0 & 0 & 0 & 0 & 0 & 1 \\
		0 & 0 & 0 & 0 & 0 & 0 & 1 & 0 & 0
	\end{pmatrix}
\end{equation}
on the state at the first maximum,
$$
	U \left(\boldsymbol{\lambda}^{(1)}\right)\, \Sigma\left(\boldsymbol{\lambda}^{(1)}\right) \, V^\dagger \left(\boldsymbol{\lambda}^{(1)}\right),
$$
where $U\left(\boldsymbol{\lambda}^{(1)}\right)$
and $V^\dagger \left(\boldsymbol{\lambda}^{(1)}\right)$
are given by eqs.~(\ref{eq:U1def}) and (\ref{eq:V1def}) and then Schmidt-decomposing the result. Using the representation (\ref{eq:Pi2_niceform}), we then find the polynomial
\begin{equation}
	{\cal P}^{(3)} (\boldsymbol{\lambda}) = P_4(\boldsymbol{\lambda})
	+ \sqrt{\lambda_0\lambda_1}\,Q_{01}(\boldsymbol{\lambda})
	+ \sqrt{\lambda_0\lambda_2}\,Q_{02}(\boldsymbol{\lambda})
	+ \sqrt{\lambda_1\lambda_2}\,Q_{12}(\boldsymbol{\lambda}),
\end{equation}
where
\begin{eqnarray}
	P_4 &=&
	0.264612\,\lambda_{0}^{4}
	+ 0.355011\,\lambda_{1}^{4}
	+ 0.300228\,\lambda_{2}^{4}
	\nonumber\\
	&+& 4.340584\,\lambda_{0}^{2}\lambda_{1}\lambda_{2}
	+ 6.394929\,\lambda_{0}\lambda_{1}^{2}\lambda_{2}
	+ 4.641264\,\lambda_{0}\lambda_{1}\lambda_{2}^{2}
	\nonumber\\
	&+& 2.454370\,\lambda_{0}^{2}\lambda_{1}^{2}
	+ 2.168776\,\lambda_{0}^{2}\lambda_{2}^{2}
	+ 2.082302\,\lambda_{1}^{2}\lambda_{2}^{2}
	\nonumber\\
	&+& 0.567507\,\lambda_{0}^{3}\lambda_{1}
	+ 0.882558\,\lambda_{0}^{3}\lambda_{2}
	+ 0.669574\,\lambda_{1}^{3}\lambda_{2}
	\nonumber\\
	&+& 0.588485\,\lambda_{0}\lambda_{2}^{3}
	+ 0.633405\,\lambda_{1}\lambda_{2}^{3}
	+ 0.377448\,\lambda_{0}\lambda_{1}^{3} ,
	\\[2mm]
	Q_{01} &=&
	- 0.001985\,\lambda_{0}^{3}
	- 0.007900\,\lambda_{1}^{3}
	+ 0.031937\,\lambda_{2}^{3}
	+ 0.037064\,\lambda_{0}\lambda_{1}\lambda_{2}
	\nonumber\\
	&& - 0.241756\,\lambda_{0}^{2}\lambda_{1}
	- 0.281329\,\lambda_{0}^{2}\lambda_{2}
	+ 0.267215\,\lambda_{1}^{2}\lambda_{2}
	\nonumber\\
	&& + 0.028893\,\lambda_{0}\lambda_{1}^{2}
	- 0.270331\,\lambda_{0}\lambda_{2}^{2}
	+ 0.542476\,\lambda_{1}\lambda_{2}^{2} ,
	\\[2mm]
	Q_{02} &=&
	-0.002437\,\lambda_{0}^{3}
	- 0.067926\,\lambda_{1}^{3}
	- 0.017591\,\lambda_{2}^{3}
	- 0.542967\,\lambda_{0}\lambda_{1}\lambda_{2}
	\nonumber\\
	&& - 0.008630\,\lambda_{0}^{2}\lambda_{1}
	+ \,0.092772\,\lambda_{0}^{2}\lambda_{2}
	- 0.322730\,\lambda_{1}^{2}\lambda_{2}
	\nonumber\\
	&& + 0.216926\,\lambda_{0}\lambda_{1}^{2}
	+ 0.225207\,\lambda_{0}\lambda_{2}^{2}
	- 0.221140\,\lambda_{1}\lambda_{2}^{2} ,
	\\[2mm]
	Q_{12} &=&
	-0.152068\,\lambda_{0}^{3}
	+ 0.009693\,\lambda_{1}^{3}
	+ 0.002967\,\lambda_{2}^{3}
	- 0.430250\,\lambda_{0}\lambda_{1}\lambda_{2}
	\nonumber\\
	&&
	- 1.309903\,\lambda_{0}^{2}\lambda_{1}
	- \,0.019484\,\lambda_{0}^{2}\lambda_{2}
	- 0.229393\,\lambda_{1}^{2}\lambda_{2} \nonumber\\
	&&
	- 0.499201\,\lambda_{0}\lambda_{1}^{2}
	- 0.004810\,\lambda_{0}\lambda_{2}^{2}
	- 0.086793\,\lambda_{1}\lambda_{2}^{2}.
\end{eqnarray}
Once again the minimal value of the polynomial ${\cal P}^{(3)}$ is $\frac{17}{81}$, found at $\boldsymbol{\lambda} = (\lambda_0^{(3)},\lambda_1^{(3)},\lambda_2^{(3)})$, whose numerical values listed in the third column of Table~\ref{tab:parameters} are obtained as the solutions to the cubic equation (\ref{eq:cubic}) with $C=\sqrt{5}/3$ and $G=9^{-1/3}$. The maximal magic near the third set of maxima is given by
\begin{equation}
	M_2^{(3)}(\lambda_0,\lambda_1,\lambda_2) =
	\max_{\substack{\text{all permutations} \\ \text{of } i, j, k}}
	\left(
	f^{(3)}(\lambda_i,\lambda_j,\lambda_k)
	\right),
	\label{eq:Mmax_case3}
\end{equation}
where the helper function is
\begin{equation}
	f^{(3)}(\lambda_0,\lambda_1,\lambda_2) = - \ln {\cal P}^{(3)} (\lambda_0,\lambda_1,\lambda_2).
\end{equation}
The function (\ref{eq:Mmax_case3}) is illustrated in the right panel of Figure~\ref{fig:fmaxfunctions}.

\begin{figure}[t]
	\centering
	\includegraphics[trim={0cm 0.0cm 0cm 0.0cm}, clip, width=0.55\linewidth]{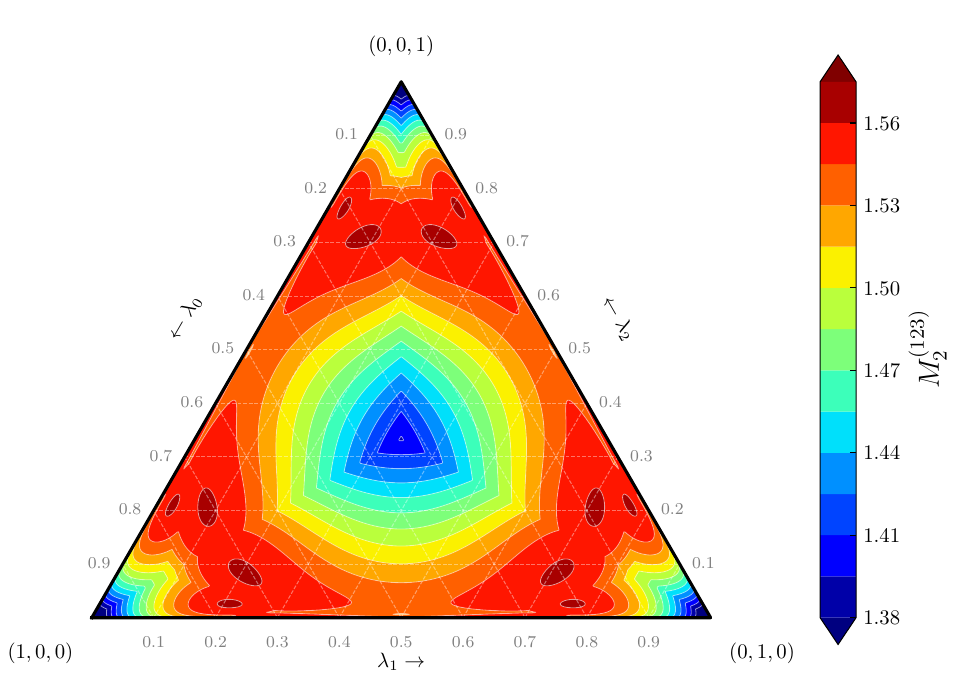}
	\hfill
	\includegraphics[width=0.43\linewidth]{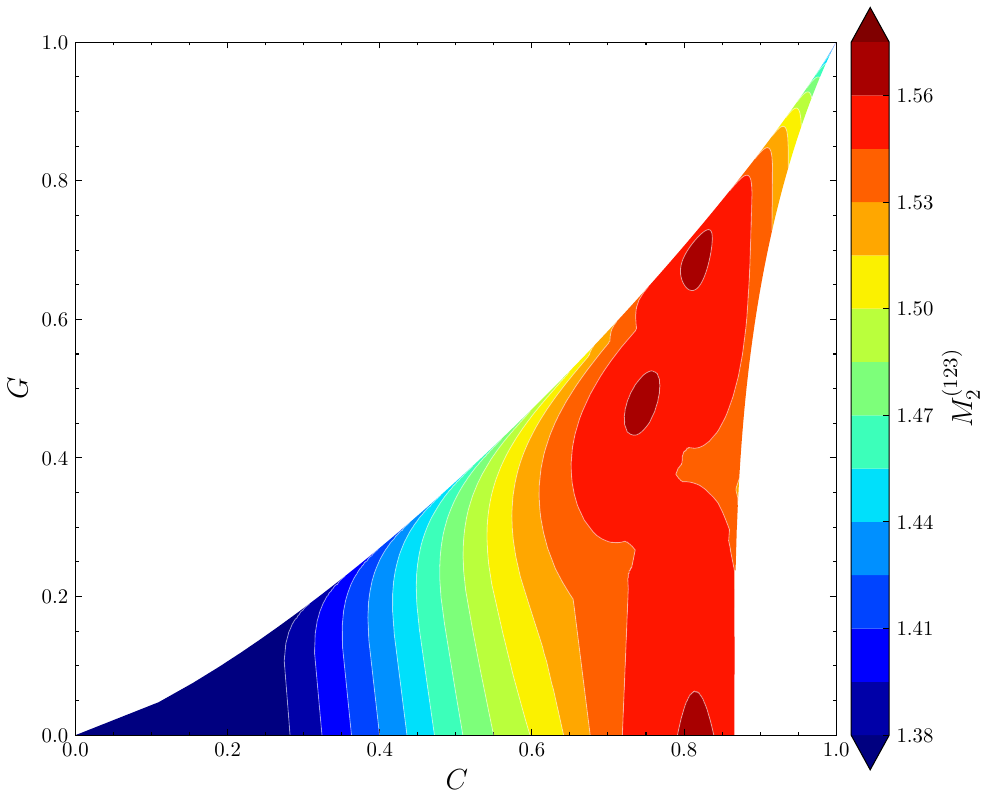}
	\caption{The same as Figures~\ref{fig:M2min} and \ref{fig:M2max}, but for the analytic function $M_2^{(123)}(\boldsymbol{\lambda})$ defined in eq.~(\ref{eq:M2_123}) as a function of $\boldsymbol{\lambda}$ (left panel) or $C$ and $G$ (right panel).
	}
	\label{fig:M2max_ana}
\end{figure}

Having found the individual functions (\ref{eq:Mmax_case1}),
(\ref{eq:Mmax_case2}) and (\ref{eq:Mmax_case3}), we can form their upper envelope
\begin{equation}
	M_2^{(123)} (\boldsymbol{\lambda})= \max_i \left( M_2^{(i)} (\boldsymbol{\lambda})\right).
	\label{eq:M2_123}
\end{equation}
The function $M_2^{(123)}(\boldsymbol{\lambda})$ defined in (\ref{eq:M2_123}) is plotted in Figure~\ref{fig:M2max_ana} in the $\boldsymbol{\lambda}$ plane (left panel) and in the $(C, G)$ plane (right panel).

\subsection{Validity Range of the Analytical Formulas for Maximal Magic}
\label{sec:validity}

In the case of two qubits, the entire Pareto boundary line of maximum magic was piecewise defined in terms of just three helper functions, $f_\text{IHG}$, $f_\text{GFE}$ and $f_\text{ED}$ (see the line IHGFED in Figure~1 in \cite{Roman:2026mcy}).
For two qutrits, the number of such helper functions that are needed to describe the full Pareto surface of maximal magic increases significantly and the derivation of each of them analytically falls beyond the scope of this paper. Instead, here we have focused on the three functions which are sufficient to describe the 18 global maxima seen in Figure~\ref{fig:M2max}. Those functions are given in eqs.~(\ref{eq:Mmax_case1}),  (\ref{eq:Mmax_case2}), and (\ref{eq:Mmax_case3}) and are plotted in Figure~\ref{fig:fmaxfunctions}. By construction, their envelope $M_2^{(123)}$, defined in eq.~(\ref{eq:M2_123}) and plotted in Figure~\ref{fig:M2max_ana}, gives the exact value for the maximal achievable magic in the neighborhood of each global maximum. As we move sufficiently far from any global maximum, we would expect that the envelope (\ref{eq:M2_123}) would cease to give the exact value of $M_{2}^{(\text{max})}$ and would only provide a lower bound on it. In this subsection, we quantitatively analyze a) the extent of the region over the full simplex where $M_2^{(123)}$ does in fact provide the exact bound; b) the remaining areas where $M_2^{(123)}$ acts strictly as a lower bound, where we shall assess the gap between the approximation (\ref{eq:M2_123}) and the true value of $M_{2}^{(\text{max})}$.
\begin{figure}
	\centering
	\includegraphics[width=0.50\linewidth]{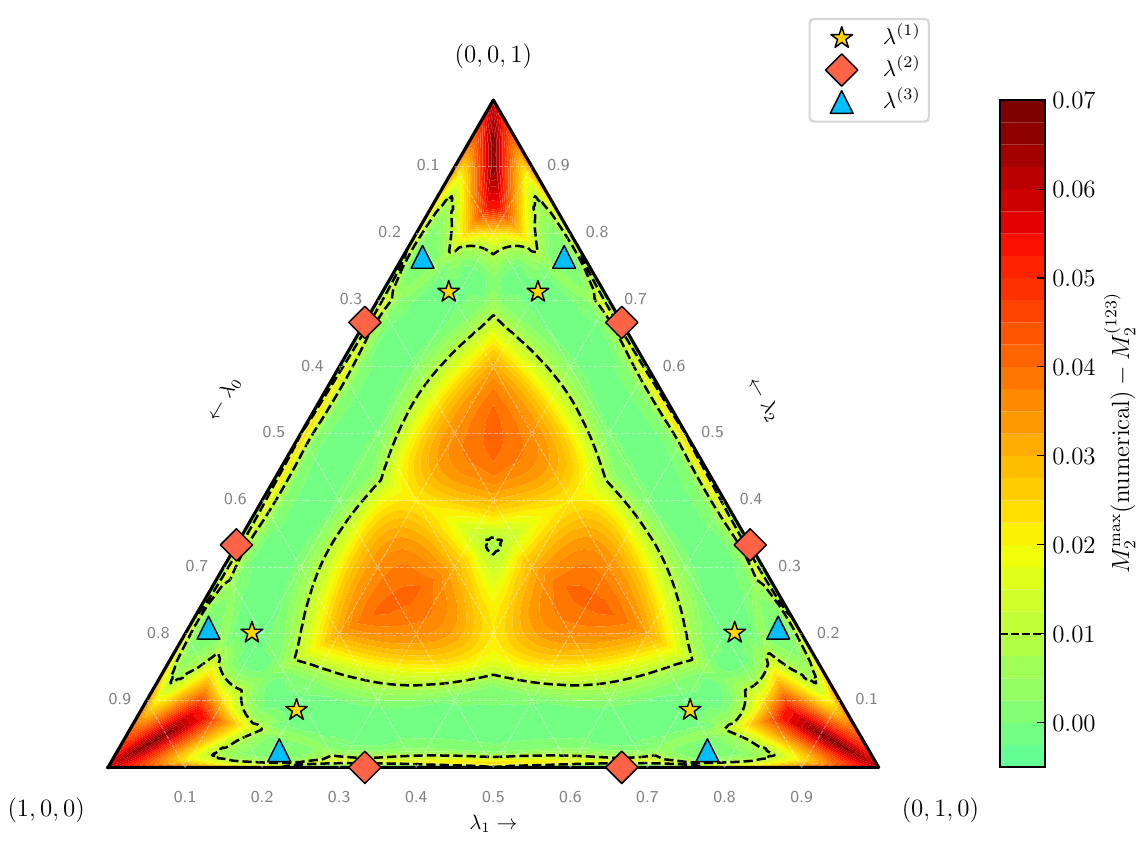}
	\includegraphics[width=0.44\linewidth]{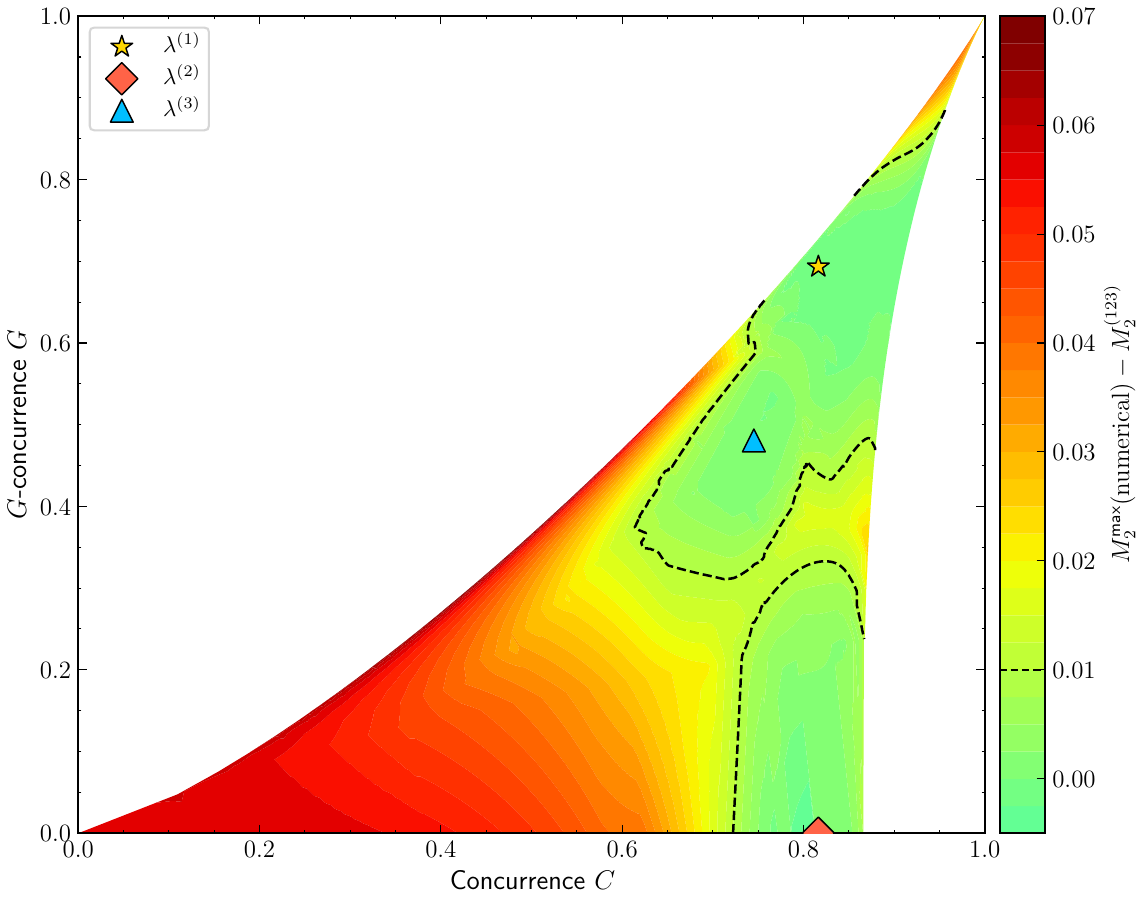}
	\caption{The difference (\ref{eq:delta_M2}) between the numerically obtained $M_2^{(\text{max})}$ in Figure~\ref{fig:M2max} and the analytical function~$M_2^{(123)}$ defined in eq.~(\ref{eq:M2_123}), plotted over the
	$(\lambda_0, \lambda_1, \lambda_2)$ simplex (left panel), or in $(C, G)$ plane (right panel).
	The dashed black line indicates the border where the difference is 0.01.
	}
	\label{fig:magic_diff_num_vs_ana}
\end{figure}
For this purpose, in Figure~\ref{fig:magic_diff_num_vs_ana} we plot the difference
\begin{equation}
	\delta M_2^{(\text{max})} (\boldsymbol{\lambda})\equiv M_2^{(\text{max})}(\boldsymbol{\lambda})- M_2^{(123)}(\boldsymbol{\lambda}),
	\label{eq:delta_M2}
\end{equation}
where the first term is the numerically obtained maximal magic $M_2^{(\text{max})}$ seen in Figure~\ref{fig:M2max} and the second term is the analytical function~$M_2^{(123)}(\boldsymbol{\lambda})$ defined in eq.~(\ref{eq:M2_123}). The dashed contour in Figure~\ref{fig:magic_diff_num_vs_ana} delineates the region where the analytical prediction $M_2^{(123)}(\boldsymbol{\lambda})$ is within $0.01$ of the true $M_2$ maximum at that $\boldsymbol{\lambda}$ location (the 0.01 threshold is less than 1\% of the global maximum value $\ln(81/17)\approx 1.56$).

Figure~\ref{fig:magic_diff_num_vs_ana} reveals large regions (colored in green) in which the analytical envelope (\ref{eq:M2_123}) does in fact provide the true maximum. To some extent, this could have been anticipated, given the similarity of Figures~\ref{fig:M2max} and \ref{fig:M2max_ana}. At the same time, there are also regions, most notably near the corners of the ternary plot, where the analytical formula $M_2^{(123)}(\boldsymbol{\lambda})$ can differ from $M_2^{(\text{max})}$ by as much as 0.07.

\begin{figure}
	\centering
	\includegraphics[width=\linewidth]{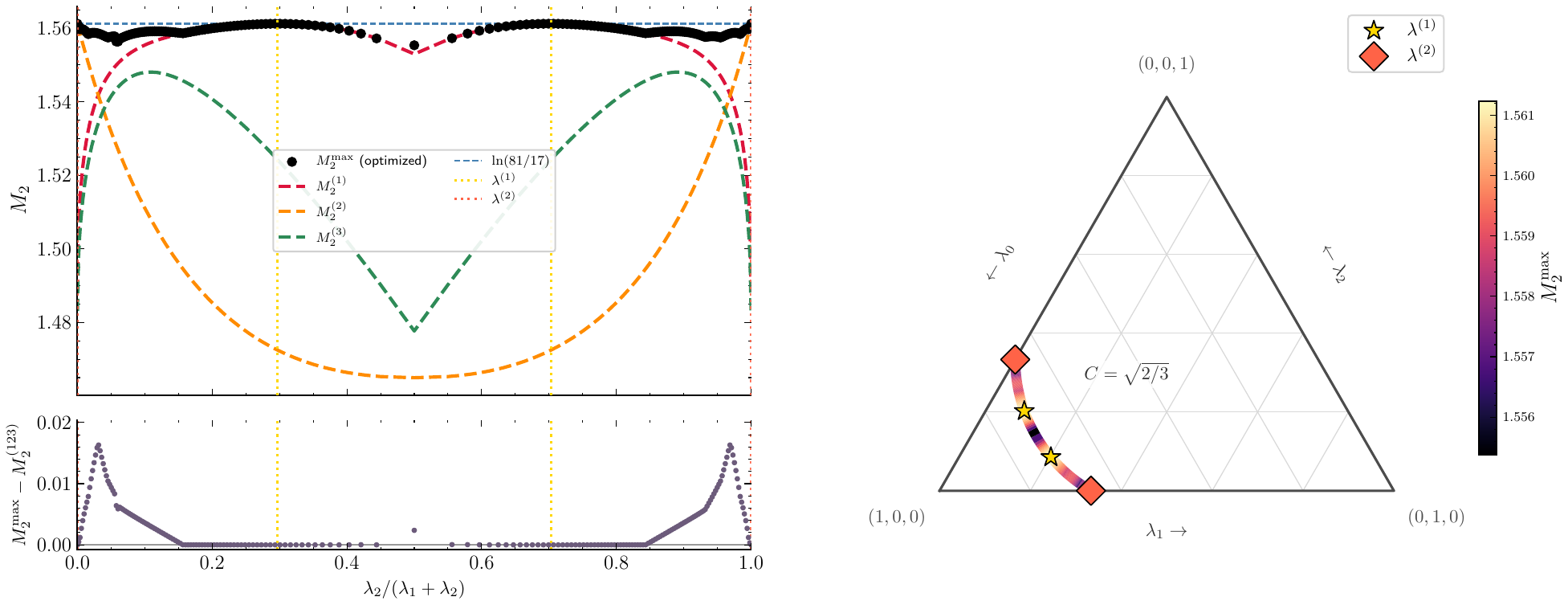}
	\caption{Maximum magic $M_2$ along a trajectory of constant $C=\sqrt{2/3}$ which passes through both maxima $\boldsymbol{\lambda}^{(1)}$ and $\boldsymbol{\lambda}^{(2)}$.
		Left upper panel: $M_2$ as a function of the trajectory parameter $\lambda_2/(\lambda_1 + \lambda_2)$.
		The black dots show the maximum magic obtained numerically and shown in Figure~\ref{fig:M2max}.
		The long-dashed lines show the functions from eqs.~(\ref{eq:Mmax_case1}), (\ref{eq:Mmax_case2}) and (\ref{eq:Mmax_case3}), while the horizontal short-dashed line marks the global maximum value $\ln(81/17)$.
		The left lower panel shows the corresponding residuals (\ref{eq:delta_M2}).
		The vertical dotted lines in the left panels and the star and diamond symbols in the right panel mark the locations of the maxima $\boldsymbol{\lambda}^{(1)}$ and $\boldsymbol{\lambda}^{(2)}$.
	}
	\label{fig:c_sqrt2_3_max_magic}
\end{figure}

\begin{figure}
	\centering
	\includegraphics[width=\linewidth]{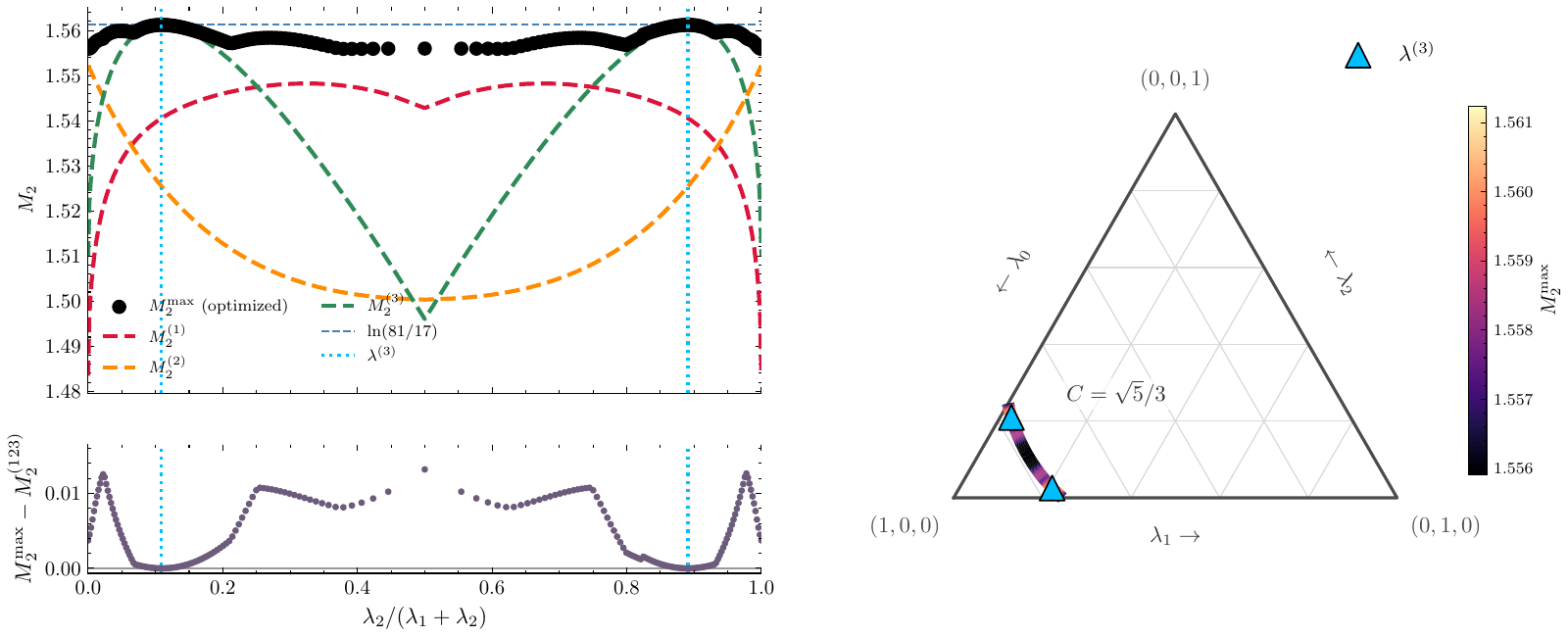}
	\caption{The same as Figure~\ref{fig:c_sqrt2_3_max_magic}, but for a constant $C=\sqrt{5}/3$ trajectory through $\boldsymbol{\lambda}^{(3)}$.}
	\label{fig:c_sqrt5_3_max_magic}
\end{figure}

\begin{figure}
	\centering
	\includegraphics[width=\linewidth]{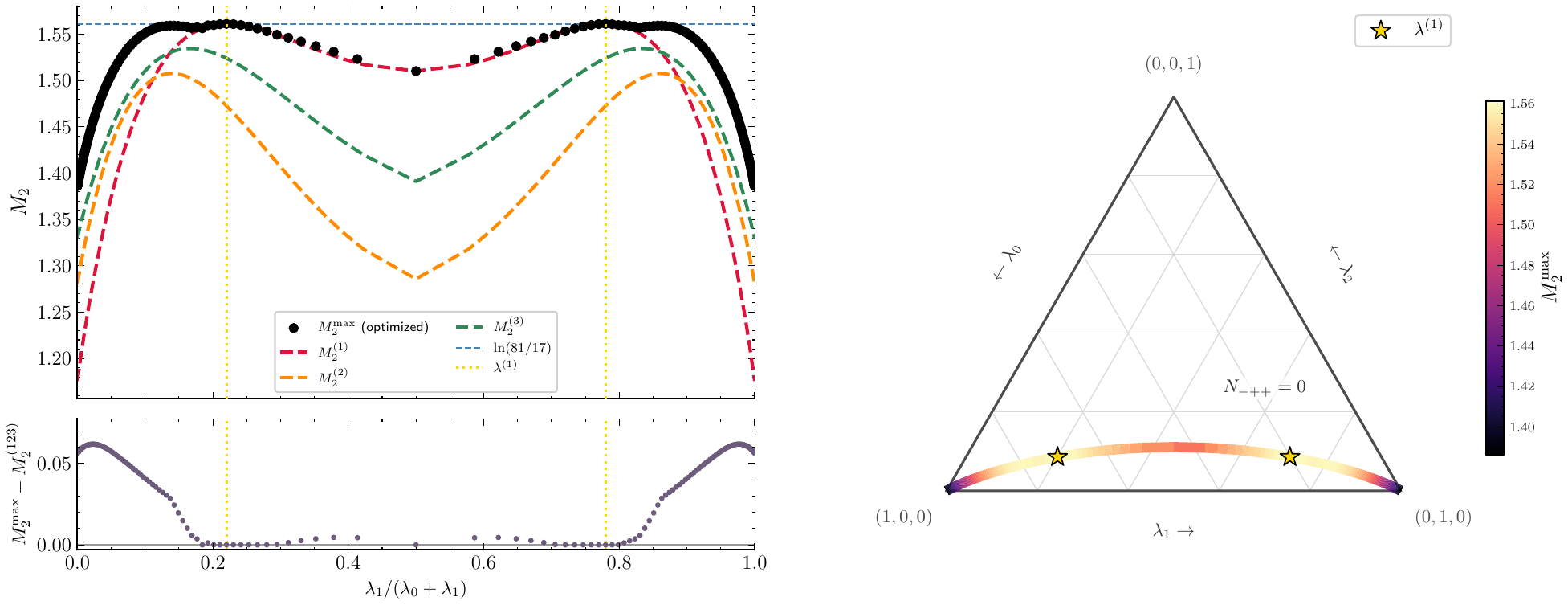}
	\caption{The same as Figures~\ref{fig:c_sqrt2_3_max_magic} and \ref{fig:c_sqrt5_3_max_magic}, but for a trajectory of constant $N_{-++}=0$, which is parametrized by the ratio $\lambda_1/(\lambda_0 + \lambda_1)$.}
	\label{fig:nmpp0_max_magic}
\end{figure}

\begin{figure}
	\centering
	\includegraphics[width=\linewidth]{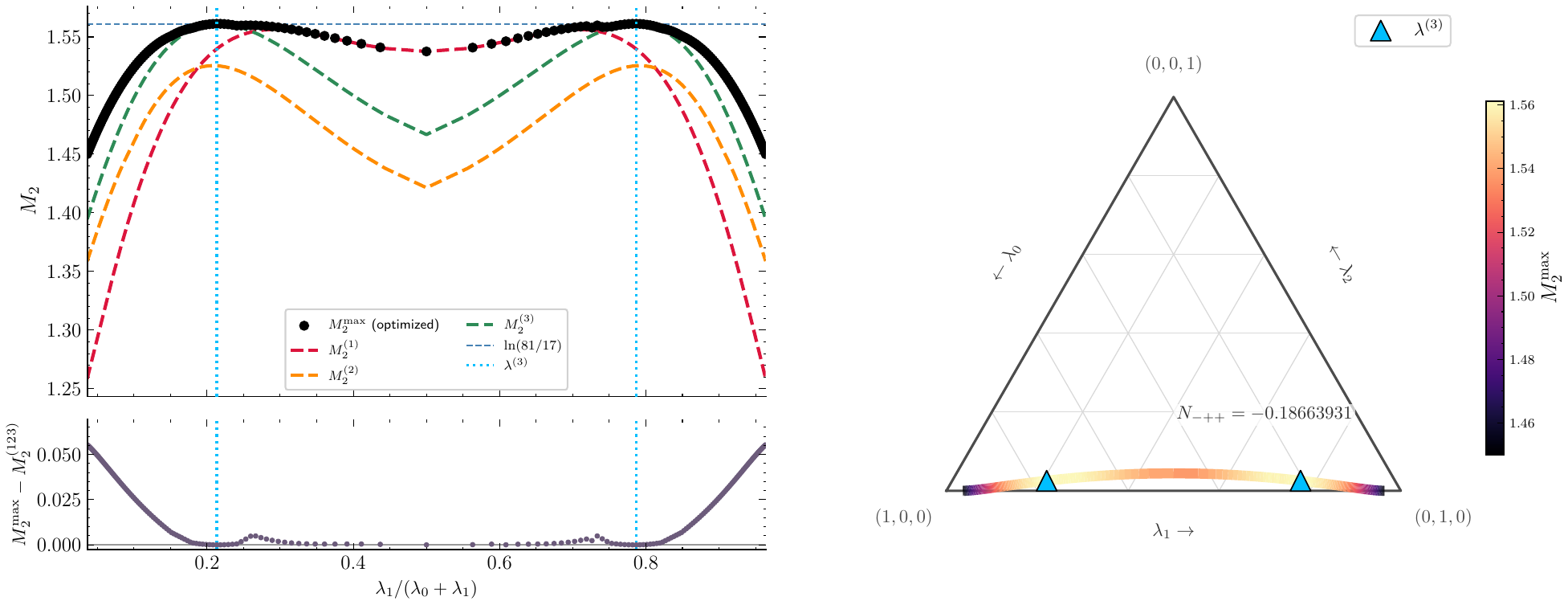}
	\caption{The same as Figure~\ref{fig:nmpp0_max_magic}, but for a trajectory of constant $N_{-++}=-0.18663931$.}
	\label{fig:nmpp_-01866_max_magic}
\end{figure}

The comparison between $M_2^{(123)}(\boldsymbol{\lambda})$ and $M_2^{(\text{max})}(\boldsymbol{\lambda})$ is explored in greater detail in Figures~\ref{fig:c_sqrt2_3_max_magic}--\ref{fig:nmpp_-01866_max_magic} along four one-dimensional cuts through the Schmidt simplex. Each cut, illustrated in the right panel of each figure, was chosen to pass straight through one or more of the global magic maxima $\boldsymbol\lambda^{(i)}$ collected in Table~\ref{tab:parameters}. Along each cut, the left upper panel displays the numerically optimized maximum $M_2^{\max}$ depicted in Figure~\ref{fig:M2max} (black dots), the three closed-form branches $M_2^{(1)}$, $M_2^{(2)}$, and $M_2^{(3)}$ (long-dashed lines) derived in the previous subsection, and the global magic ceiling $\ln(81/17)$ (horizontal short-dashed line). The lower left panel in each figure shows the residuals $\delta M_2^{(\text{max})}$ defined in eq.~(\ref{eq:delta_M2}).

The horizontal axes in the left panels of Figures~\ref{fig:c_sqrt2_3_max_magic}--\ref{fig:nmpp_-01866_max_magic} are parametrized using specific rational expressions of the Schmidt weights $\lambda_i$.
These ratios are chosen to smoothly represent the trajectories across the simplex:
\begin{itemize}
	\item For the trajectories of constant $C=\sqrt{2/3}$ and $C=\sqrt{5}/3$ (Figures~\ref{fig:c_sqrt2_3_max_magic} and~\ref{fig:c_sqrt5_3_max_magic}), the axis parameter is $\lambda_2 / (\lambda_1 + \lambda_2)$ and it varies between 0 and 1.
	\item For the trajectories of constant $N_{-++}=1/3$ and $N_{-++}=-0.18663931$ (Figures~\ref{fig:nmpp0_max_magic} and~\ref{fig:nmpp_-01866_max_magic}), the axis is parameterized by $\lambda_1 / (\lambda_0 + \lambda_1)$ and also varies between 0 and 1.
\end{itemize}

Figures~\ref{fig:c_sqrt2_3_max_magic}--\ref{fig:nmpp_-01866_max_magic} confirm that each closed-form expression $M_2^{(i)}$ reproduces the magic exactly in the neighborhood of the global maximum from which it was derived.
Away from that maximum, it might fall below the true $M_2^{(\text{max})}$.
The pointwise upper envelope, $M_2^{(123)}$, is therefore, by construction, a lower bound on the optimized magic: $M_2^{(123)}(\lambda)\le M_2^{\max}(\lambda)$ at every point of the simplex, with equality wherever one of the three branches is exact.
Along the four cuts of Figures~\ref{fig:c_sqrt2_3_max_magic}--\ref{fig:nmpp_-01866_max_magic}, the envelope and the numerical optimum agree very well, to better than $1\%$ along the constant $C$ cuts in Figures \ref{fig:c_sqrt2_3_max_magic} and \ref{fig:c_sqrt5_3_max_magic} and to better than $5\%$ over the constant $N$ cuts in Figures~\ref{fig:nmpp0_max_magic} and \ref{fig:nmpp_-01866_max_magic}.

\section{Generalization Beyond Qutrits}
\label{sec:generalization}

A particularly elegant class of non-stabilizer states arises from the orbit of the Weyl-Heisenberg (WH) group acting on a single fiducial vector.
For $d$-qudits, the WH group is generated by the generalized Pauli shift and clock operators (\ref{eq:shiftclock}), whose products (\ref{eq:Padef}) furnish $d^4$ displacement operators labeled by $\boldsymbol{a} = (a_1, a_2, b_1, b_2)$.
A fiducial vector $|\psi\rangle$ is said to be WH-covariant for mutually unbiased bases (MUBs) if its orbit $\{P_{\boldsymbol{a}}|\psi\rangle\}$ organizes into $d^2+1$ orthonormal bases of $d^2$ vectors each, with vectors from any two distinct bases overlapping with the maximally unbiased magnitude $1/d$~\cite{Blanchfield:2014ogi,Feng:2024whz}.
Following on the initial observation of \cite{Liu:2025frx} for the qubit case, we have checked that the states discussed above which maximize $M_2$ are WH-covariant MUB fiducial states~\cite{Blanchfield:2014ogi,Feng:2024whz, Ivonovic:1981pvj, WOOTTERS1989363}, generalizing the analogous result connecting maximal magic to symmetric informationally complete (SIC) fiducial states~\cite{Cuffaro:2024wet}. 

We can use this observation to generalize the initial result (\ref{eq:Cuffaro_bound}) to the case of any prime $d$.
\begin{equation}
	\max(M_2) = \ln\frac{d^2}{1+\frac{d^2-1}{d^2}} = \ln \frac{d^4}{2d^2-1}.
	\label{eq:bound}
\end{equation}
This formula can be understood as follows. For each $d$, there are $d^4$ displacement operators. Among those, these are $d^2$ operators that map $|\psi\rangle$ to states within its own MUB:
the identity operator maps $|\psi\rangle$ to itself and therefore contributes 1 to the sum in (\ref{eq:purity}), while the other $d^2-1$ operators map $|\psi\rangle$ to the remaining $d^2-1$ basis vectors in the same MUB, which are orthogonal to $|\psi\rangle$ by definition and hence contribute zero to the purity sum.
Finally, the remaining $d^4-d^2$ operators map $|\psi\rangle$ to states in the other $d^2$
MUBs; by the definition of mutual unbiasedness, they have equal contributions given by $\frac{1}{d^4}$. As a result, the sum entering the definition of purity is $1+(d^2-1)\cdot 0 + (d^4-d^2)\cdot \frac{1}{d^4}$, which is precisely the denominator seen above.

\subsection{A System of Two Ququints}
\label{sec:ququints}

For $d=2$, the bound (\ref{eq:bound}) reproduces the previous result of $\ln (16/7)$ for qubits derived in \cite{Liu:2025frx}. For $d=3$, it gives the answer of $\ln (81/17)$ for qutrits found in Section~\ref{sec:maxmagic} above. In order to test if this pattern continues, we now consider the case of two ququints ($d=5$). The results are summarized in Table~\ref{tab:ququints}.
\begin{table}[t]
	\caption{\label{tab:ququints}
		The same as Table~\ref{tab:parameters} but for the case of a system of two ququints. There are six different degenerate maxima of the magic.}
	\begin{ruledtabular}
		\begin{tabular}{ccccccc}
			Var                  & Max 1                                       & Max 2                        & Max 3                      & Max 4                      & Max 5                        & Max 6                \\
			\colrule
			$\lambda_0$          & $\dfrac{2}{5}$
			                     & $\dfrac{3(5+\sqrt5)}{50}$
			                     & $0.500137...$
			                     & $0.542394...$
			                     & $\dfrac{10+3\sqrt5+\sqrt{75+30\sqrt5}}{50}$
			                     & $\dfrac{3+\sqrt5}{10}$                                                                                                                                                                     \\[2mm]
			$\lambda_1$          & $\dfrac{2}{5}$
			                     & $\dfrac{3(5+\sqrt5)}{50}$
			                     & $0.366872...$
			                     & $0.275277...$
			                     & $\dfrac{3(5-\sqrt5)}{50}$
			                     & $\dfrac{1}{5}$                                                                                                                                                                             \\[2mm]
			$\lambda_2$          & $\dfrac{1}{5}$
			                     & $\dfrac{10-3\sqrt5+\sqrt{75-30\sqrt5}}{50}$
			                     & $0.063283...$
			                     & $\dfrac{5+\sqrt5}{50}$
			                     & $\dfrac{3(5-\sqrt5)}{50}$
			                     & $\dfrac{1}{5}$                                                                                                                                                                             \\[2mm]
			$\lambda_3$          & $0$
			                     & $\dfrac{10-3\sqrt5-\sqrt{75-30\sqrt5}}{50}$
			                     & $\dfrac{5-\sqrt5}{50}$
			                     & $0.032582...$
			                     & $\dfrac{10+3\sqrt5-\sqrt{75+30\sqrt5}}{50}$
			                     & $\dfrac{3-\sqrt5}{10}$                                                                                                                                                                     \\[2mm]
			$\lambda_4$          & $0$                                         & $0$                          & $0.014430...$              & $0.005025...$              & $0$                          & $0$                  \\
			[1mm]
			\hline
			$e_2$                & $\dfrac{8}{25}$                             & $\dfrac{38}{125}$            & $\dfrac{38}{125}$          & $\dfrac{38}{125}$          & $\dfrac{38}{125}$            & $\dfrac{8}{25}$      \\[1mm]
			$e_3$                & $\dfrac{4}{125}$                            & $\dfrac{3(13-\sqrt5)}{1250}$ & $\dfrac{95-2\sqrt5}{3125}$ & $\dfrac{95+2\sqrt5}{3125}$ & $\dfrac{3(13+\sqrt5)}{1250}$ & $\dfrac{1}{25}$      \\[1mm]
			$e_4$                & $0$                                         & $\dfrac{9(3-\sqrt5)}{31250}$ & $\dfrac{29+\sqrt5}{31250}$ & $\dfrac{29-\sqrt5}{31250}$ & $\dfrac{9(3+\sqrt5)}{31250}$ & $\dfrac{1}{625}$     \\[1mm]
			$e_5$                & $0$                                         & $0$                          & $\dfrac{5+\sqrt5}{781250}$ & $\dfrac{5-\sqrt5}{781250}$ & $0$                          & $0$                  \\
			[2mm]
			\colrule
			$M_2^{(\text{max})}$ & $\ln\dfrac{625}{49}$                        & $\ln\dfrac{625}{49}$         & $\ln\dfrac{625}{49}$       & $\ln\dfrac{625}{49}$       & $\ln\dfrac{625}{49}$         & $\ln\dfrac{625}{49}$ \\
		\end{tabular}
	\end{ruledtabular}
\end{table}
In analogy to (\ref{eq:cubic}), the Schmidt coefficients satisfy the quintic equation
\begin{equation}
	\lambda^5 - \lambda^4 + e_2 \lambda^3 - e_3 \lambda^2 + e_4 \lambda - e_5 =0,
\end{equation}
where
\begin{eqnarray}
	e_2 &\equiv& \sum_{i<j} \lambda_i\lambda_j, \\
	e_3 &=& \sum_{i<j<k} \lambda_i\lambda_j\lambda_k, \\
	e_4 &=& \sum_{i<j<k<l} \lambda_i\lambda_j\lambda_k \lambda_l, \\
	e_5 &=& \lambda_0 \lambda_1 \lambda_2 \lambda_3 \lambda_4.
\end{eqnarray}
The values for $e_2$, $e_3$,  $e_4$ and  $e_5$ at each maximum are given in the middle section of the table. Most of the Schmidt coefficients in Table~\ref{tab:ququints} are given analytically with simple expressions. In the case of Maxima 3 and 4, four of the Schmidt coefficients are obtained from a quartic equation and are listed numerically. In all six cases, the maximal magic is found to be $\ln (625/49)$, again in agreement with the prediction of eq.~(\ref{eq:bound}). This suggests that eq.~(\ref{eq:bound}) perhaps applies to the case of any prime $d$ beyond $d=5$, but we have no formal proof.

\section{Conclusions}
\label{sec:conclusions}

In this work, we extended the program of mapping the Pareto frontiers of magic and entanglement from two qubits to two qutrits.
The qualitative novelty of the qutrit case is that the Schmidt spectrum carries two independent entanglement parameters instead of just one, so that the frontiers are no longer one-dimensional curves, but become two-dimensional surfaces over the Schmidt simplex.
We characterized both boundaries.
For the lower frontier, we rederived the minimal magic from Ref.~\cite{Busoni:2026lvp} and recast it in eq.~(\ref{eq:M2min_CG}) as a compact function of the $I$-concurrence and negativity, $M_2^{(\text{min})}(C,N)$, which has a maximum value of $\ln 2$.
For the upper frontier, we found that the largest attainable magic is $\ln(81/17)\approx 1.561$, achieved at eighteen types of states that fall into three permutation-inequivalent families of six. The states from the three families are related to each other through Clifford operations.
For each family, we constructed a closed-form expression for the maximal magic which is valid in the neighborhood of the corresponding maximum.

Our value of $\ln(81/17)$ tightens the earlier upper bound of $\ln 5\approx 1.609$ of Refs.~\cite{Wang:2023uog,Cuffaro:2024wet} and improves the numerical estimate of Ref.~\cite{Chernyshev:2024pqy}.
More importantly, the maximizing states turn out to be Weyl--Heisenberg-covariant fiducial states for mutually unbiased bases.
This observation allowed us to conjecture a closed-form maximal magic for two qudits of any prime dimension $d$, eq.~(\ref{eq:bound}), which reproduces the known qubit value from \cite{Liu:2025frx} and the qutrit and ququint values derived here in Sections~\ref{sec:maxmagic} and \ref{sec:ququints}, respectively.

This analysis also identifies remaining gaps in knowledge.
As shown in Section~\ref{sec:validity}, the upper envelope $M_2^{(123)}$ of our three analytic branches is, by construction, a genuine lower bound on the true maximal magic, and is exact in extended regions around the maxima and along the edges connecting them.
It is not, however, exact everywhere: a structured, three-fold-symmetric residual survives in a central region of the simplex and near the corners, where the gap reaches a few percent of $\ln(81/17)$.
The existence of these residuals shows that the three families do not exhaust all the states which provide locally maximal magic.
Away from the maxima the magic-optimal local unitaries drift to new configurations, each of which defines its own analytic branch through the construction used here.
We do not attempt to write all of them down.
While our construction outlined in this paper could in principle be repeated at any Clifford-optimal configuration, the number of distinct branches needed to reproduce $M_2^{(\text{max})}(\boldsymbol{\lambda})$ pointwise appears to proliferate as one moves away from the maxima, so that a complete piecewise-analytic tiling of the simplex would be unwieldy and of limited additional value.
We therefore regard the three branches as exact local descriptions anchored at the global maxima, sufficient to establish the maximal value $\ln(81/17)$ and to capture the magic in the most relevant regions of the simplex.
The surface as a whole is most economically characterized numerically, as in Figure~\ref{fig:M2max}.
Determining the minimal set of branches required for a prescribed accuracy, or finding a more compact closed form that bypasses the branch-by-branch construction altogether, remains an interesting open problem which is beyond the scope of this paper.

Beyond two qutrits, several directions invite further study.
The most immediate is a proof of the conjectured prime-$d$ bound~(\ref{eq:bound}) and an understanding of how the number and structure of the maxima grow with $d$, as already hinted at by the richer pattern of six maxima found for the ququint case.
The connection between maximal-magic states and Weyl--Heisenberg-covariant MUB fiducials ties the resource-theoretic question of maximal nonstabilizerness to long-standing existence problems in finite quantum geometry~\cite{doi:10.1142/S0219749911006776, Appleby:2005plb}.
Finally, it would be worthwhile to extend the present pure-state analysis to mixed states and to other magic monotones, and to ask whether the frontiers derived here carry operational consequences for the cost of classically simulating qutrit-based quantum circuits.

\begin{acknowledgments}
	We thank H.~Lamm and P.~Vander Griend for useful discussions.
	The work of AR, MK and KTM is supported in part by the Shelby Endowment for Distinguished Faculty at the University of Alabama. The work of AR and KM is supported in part by Fermilab via Subcontract 731293, in support of DOE Award No.\ DE-SCL0000090 ``HEP AmSC IDA Pilot: Knowledge Extraction'' and DOE Award No.\ DE-SCL0000152 ``USQCD AmSC Infrastructure Provision''. The work of KM is supported in part by the U.S. Department of Energy (DOE) under Award No. DE-SC0026347.
\end{acknowledgments}

\appendix
\section{Numerical optimization}
\label{app:numerics}

We evaluate the magic directly from its definition.
For a given state $\ket{\psi}$ we compute all $81$ Pauli expectation values $\bra{\psi}P_{\mathbf a}\ket{\psi}$ of eq.~(\ref{eq:Padef}) and assemble the stabilizer purity $\Pi_2$ and the magic $M_2=-\ln\Pi_2$ through eqs.~(\ref{eq:purity}) and~(\ref{eq:M2_def}).
The expectation values are obtained by contracting the (reshaped) density matrix $\ket{\psi}\!\bra{\psi}$ against the single-qutrit Heisenberg--Weyl operators one qutrit at a time, which avoids ever forming the full $81$-operator basis explicitly and renders $M_2$ a smooth, automatically differentiable function of the state.
All computations are performed in double precision.

To locate the extremal magic at a fixed Schmidt spectrum $\boldsymbol\lambda$ we parametrize the local unitaries through the $\mathfrak{u}(3)$ Lie algebra,
\begin{equation}
	U_X(\boldsymbol\varphi_X) = \exp\!\Big(i\sum_{a=0}^{8}\varphi_{X,a}\,T_a\Big),
	\qquad X=A,B,
	\label{eq:U_expmap}
\end{equation}
where $T_0=\mathds{1}$ and $T_1,\dots,T_8$ are the eight Gell-Mann matrices, so that each unitary is encoded by a vector of nine real parameters $\boldsymbol\varphi_X\in\mathbb R^9$ and the optimization is unconstrained.
This parametrization is deliberately over-complete: the generator $T_0=\mathds{1}$ generates only an overall phase, which cancels in $U_A\otimes U_B$ and hence does not affect the magic.

Substituting eq.~(\ref{eq:U_expmap}) into the Schmidt decomposition~(\ref{eq:Schmidt_decomposition_LU}) expresses the state, and therefore $M_2$, as a differentiable function of the $18$ real parameters $(\boldsymbol\varphi_A,\boldsymbol\varphi_B)$ at fixed $\boldsymbol\lambda$.
We extremize it with the Adam optimizer, computing gradients by automatic differentiation.
The same routine serves both Pareto frontiers: minimizing the objective yields the lower frontier $M_2^{(\mathrm{min})}$ of Section~\ref{sec:minmagic}, while maximizing it (equivalently, minimizing $\Pi_2$) yields the upper frontier $M_2^{(\mathrm{max})}$ of Section~\ref{sec:maxmagic}.
Since the landscape is non-convex and harbors many local optima, for each $\boldsymbol\lambda$ we repeat the optimization from several independent random initializations (250--600 steps each) and keep the best result.
Scanning $\boldsymbol\lambda$ across the Schmidt simplex in this way produces the numerical surface of Figure~\ref{fig:M2max}.

\begin{figure}[t]
	\centering
	\includegraphics[width=\textwidth]{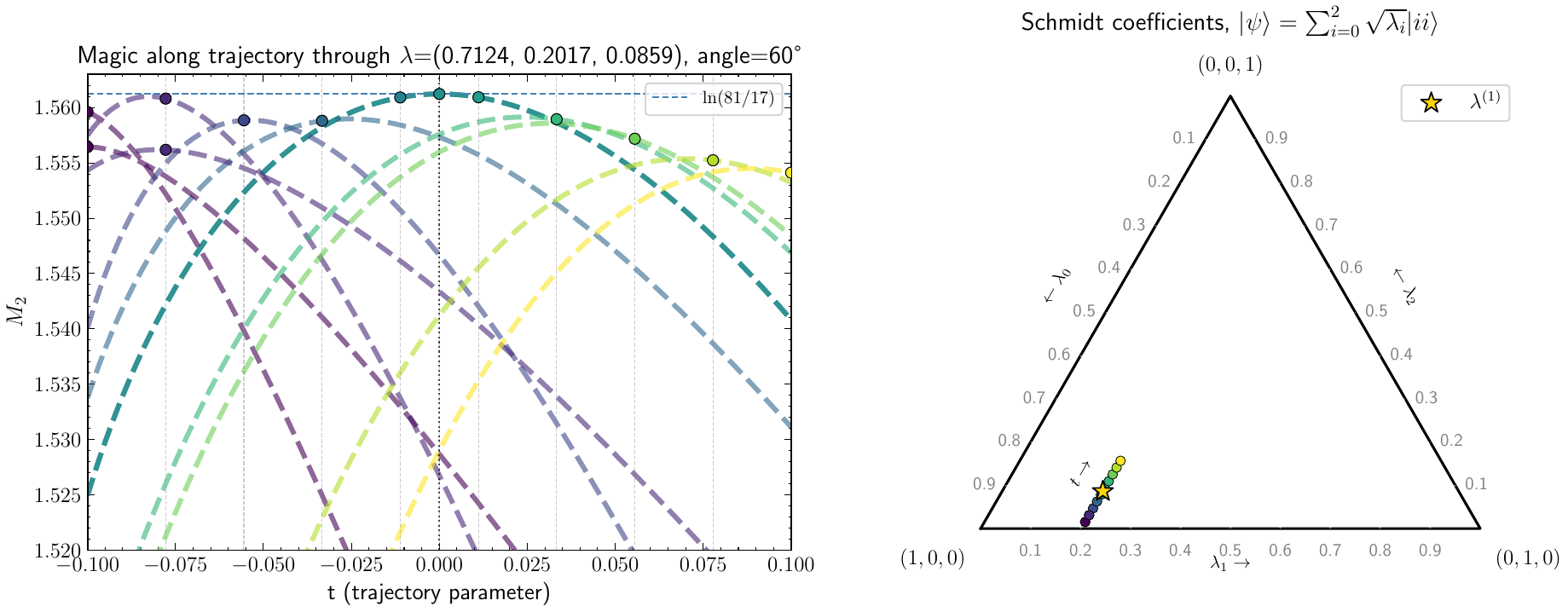}
	\caption{Construction of the maximal magic as the upper envelope of
		fixed-unitary curves.
		\textbf{Right:} a one-parameter trajectory through the Schmidt simplex,
		passing through one of the global maxima; colored markers indicate the
		sampled spectra $\boldsymbol\lambda$.
		\textbf{Left:} for each sampled spectrum the local unitaries are optimized to
		maximize $M_2$ (markers); holding each optimal pair of unitaries fixed and
		varying $\boldsymbol\lambda$ along the trajectory traces out the dashed
		curves. The numerically optimized maximal magic (solid) is the upper envelope
		of these fixed-unitary curves. The horizontal dashed line marks the global
		ceiling $\ln(81/17)$.
	}
	\label{fig:lambda_0_trajectory}
\end{figure}

The numerical optimum at a given $\boldsymbol\lambda$ returns not only the value of the magic but also the magic-optimal local unitaries, and this is the bridge to the closed-form results of Section~\ref{sec:maxmagic}. Freezing the optimal
$U_A$ and $U_B$ and letting the Schmidt spectrum vary turns $\Pi_2$ into an exact polynomial in the amplitudes $\sqrt{\lambda_i}$ of the form given in eqs.~(\ref{eq:Pi2_polynomial}) and~(\ref{eq:Pi2_niceform}).
Geometrically, a frozen pair of unitaries traces the magic of one fixed Clifford-optimal configuration as $\boldsymbol\lambda$ is varied, and the true maximal magic is recovered as the upper envelope of all such curves, as illustrated in the left panel of Figure~\ref{fig:lambda_0_trajectory}.
Note that in this figure we deliberately did not reduce to the single best magic at each $\boldsymbol\lambda$.
Instead, we retained every distinct solution found across the random seeds.
This is visible at small $t$, where
we sometimes found local maxima (at $t=-0.1$ and $t=-0.078$).

We subjected the pipeline to several independent checks.
The magic routine returns $M_2=0$ for stabilizer states (computational-basis states and Heisenberg--Weyl eigenstates) and strictly positive values for generic states.
The numerically optimized magic agrees with the frozen-unitary symbolic expressions wherever the latter are exact.
The three degenerate maxima all return $M_2=\ln(81/17)$, i.e.\ the minimal purity reaches $\Pi_2 = 17/81$ at each, and they are mapped onto one another by Clifford operations, consistent with the Clifford invariance of $M_2$.
Finally, the numerically found lower frontier reproduces the closed-form $M_2^{(\mathrm{min})}$ of eq.~(\ref{eq:M2min_lambdas}).

\bibliography{mybib}

\end{document}